\documentclass[twocolumn]{aastex62}

\newcommand\harm{\textit{HARM${^2}$}}
\newcommand\orion{\texttt{ORION}}
\def\subvir{\texttt{SubVir}}
\def\vir{\texttt{Vir}}
\def\pap{\texttt{Paper~I}}

\usepackage[normalem]{ulem}

\newcommand{\blue}[1]{{\textcolor{blue}{#1}}}
\newcommand{\black}[1]{{\textcolor{black}{#1}}}

\newcommand{\add}[1]{{\black{#1}}}

\newcommand{\addd}[1]{{\black{#1}}}

\received{February 25, 2019}
\accepted{\today}
\submitjournal{ApJ}

\shorttitle{Massive Star Formation from Subvirial and Virialized Collapse}
\shortauthors{Rosen et al.}

\begin{document}

\title{Massive Star Formation via the Collapse of Subvirial and Virialized Turbulent Massive Cores}

\correspondingauthor{Anna Rosen}
\email{anna.rosen@cfa.harvard.edu}

\author[0000-0003-4423-0660]{Anna L. Rosen}
\altaffiliation{NASA Einstein Fellow}
\altaffiliation{ITC Fellow}
\affiliation{Center for Astrophysics $|$ Harvard \& Smithsonian, 60 Garden St, Cambridge, MA 02138, USA}

\author{Pak Shing Li}
\affiliation{Astronomy Department, University of California, Berkeley, CA 94720, USA}

\author{Qizhou Zhang}
\affiliation{Center for Astrophysics $|$ Harvard \& Smithsonian, 60 Garden St, Cambridge, MA 02138, USA}

\author{Blakesley Burkhart}
\affiliation{Center for Computational Astrophysics, Flatiron Institute,162 Fifth Avenue, New York, NY 10010, USA}
\affiliation{Department of Physics and Astronomy, Rutgers, The State University of New Jersey,136 Frelinghuysen Rd, Piscataway, NJ 08854, USA}

\begin{abstract}

Similar to their low-mass counterparts, massive stars likely form via the collapse of pre-stellar molecular cores. Recent observations suggest that most massive cores are subvirial (i.e., not supported by turbulence) and therefore are likely unstable to gravitational collapse. Here we perform radiation hydrodynamic simulations to follow the collapse of turbulent massive pre-stellar cores with subvirial and virialized initial conditions to explore how their dynamic state affects the formation of massive stars and core fragmentation into companion stars. We find that subvirial cores undergo rapid monolithic collapse resulting in \addd{higher accretion rates at early times as compared to the collapse of virialized cores that have the same physical properties.}
In contrast, we find that virialized cores undergo a slower, gradual collapse and \add{significant} turbulent fragmentation at early times resulting in numerous companion stars. In the absence of strong magnetic fields \add{and protostellar outflows} we find that the faster growth rate of massive stars that are born out of subvirial cores leads to an increase in the radiative heating of the core thereby further suppressing fragmentation at early times when turbulent fragmentation occurs for virialized cores. \addd{Regardless of initial condition, we find that the massive accretion disks that form around massive stars dominant the accretion flow onto the star at late times and eventually become gravitationally unstable and fragment to form companion stars at late times.} 
\end{abstract}

\keywords{methods: numerical --- stars: formation --- stars: massive --- turbulence}

\section{Introduction} 
\label{sec:intro}
Massive stars play an essential role in the Universe. Their explosive deaths produce most of the heavy elements, enriching the interstellar medium (ISM) and future generations of stars. They are rare, yet the energy and momentum they inject into the ISM dwarfs the contribution of their more numerous low-mass cousins. This stellar feedback may set an upper limit on stellar masses, thereby affecting elemental abundances in the Universe. While the universal importance of massive stars is well understood the initial conditions, accretion history, and time span of their formation remains debated.

Massive stars form in dense ($\gtrsim10^{4}-10^{\add{6}} \;  \rm{cm^{-3}}$), cold ($\sim$10~K) turbulent gas within giant molecular clouds and giant massive filaments \citep[\add{e.g.,}][]{Smith2009a, Zhang2015a, Williams2018a}. These condensations, commonly called clumps, have masses ranging from a $\sim$few tens to $10^5 \; M_{\rm \odot}$ and sizes of $\simeq 0.5 - 2$ pc in the Milky Way \add{\citep[e.g.,][]{Urquhart2018a}}. The most massive (i.e., those with $M \gtrsim 10^3 \; M_{\rm \odot}$) and quiescent of these clumps have high surface densities with $\Sigma \gtrsim 0.05 \; \rm{g \;cm^{-2}}$ and are likely the sites of massive star and protocluster formation \citep{Traficante2015a}. 

These clumps host massive pre-stellar cores with typical sizes of 0.1~pc, which is a requirement for the Turbulent Core model for massive star formation \citep{McKee2003a}. In this model, massive stars form in a similar fashion to their low-mass counterparts via the monolithic collapse of massive pre-stellar cores that are supported by turbulence rather than thermal motions \citep{McKee2003a}. The stability of these cores can be described by their virial parameter given by $\alpha_{\rm vir} = 5 \sigma_{\rm 1D}^2 R_{\rm c}/G M_{\rm c}$ where $\sigma_{1D}$ is the core's 1D velocity dispersion and $M_{\rm c}$ and $R_{\rm c}$ are the core mass and radius, respectively \citep{Bertoldi1992a}. Neglecting external pressure and magnetic fields, cores with $\alpha_{\rm vir} \gtrsim 1$ are stable against gravitational collapse whereas those with $\alpha_{\rm vir} < 1$ are subvirial and unstable to collapse. 

The Turbulent Core model requires that massive pre-stellar cores are in approximate virial equilibrium (i.e., $\alpha_{\rm vir} \simeq 1$). These cores then become marginally unstable to collapse to form a massive star or massive multiple system. The resulting formation time scale is several times the core free-fall time scale \add{($t_{\rm ff}  \la 10^5$ yr)} and the high degree of turbulence causes clumping, resulting in high accretion rates ($\dot{M}_{\rm acc} \sim 10^{-4}-10^{-3} \; \rm{M_{\rm \odot} \; s^{-1}}$) that can overcome the radiation pressure associated with the star's large luminosity \citep{McKee2003a}. In agreement with this picture, observations show that massive cores live in highly pressurized environments and have non-thermal turbulent motions that dominate over thermal motions \citep[\add{e.g.,}][]{Plume1997a, Tan2013a, ZhangTan2015a, Zhang2015a, LiuTan2018a}. \add{Additionally, due to the core's high angular momentum content, an optically thick accretion disk forms around the accreting massive star as the core collapses and delivers material at high rates via gravitational torques to the star \citep{Yorke2002a}.}

On smaller scales, massive clumps and cores fragment into pre-stellar cores via turbulent fragmentation that have masses larger then the mass and length scale dictated by thermal Jeans fragmentation \citep{Jeans1902a, Padoan2001a, Padoan2002a, Offner2009a, Zhang2009a, Wang2014a}. \add{Additionally, observations of massive protostellar cores show that they further undergo thermal Jeans fragmentation \citep[e.g.,][]{Palau2015a, Beuther2019a}}. \add{Such fragments may be the pre-cursers of low-mass pre-stellar cores that can} accrete from the surrounding unbound gas to form massive stars as described by the Competitive Accretion model \citep{Bonnell2001a, Bonnell2006a}. This model posits that a low-mass protostellar seed will accrete unbound gas within the clump \add{as determined by its tidal limits and when it becomes massive enough it will then accrete} at the Bondi-Hoyle accretion rate, $\dot{M}_{\rm \star, \, BH} \propto v^{-3}$ where $v$ is the relative velocity of the gas. This model achieves high accretion rates onto the proto\add{star} under subvirial initial conditions, in contrast to the virialized conditions of the Turbulent Core model, since the gas velocity dispersion, and hence the level of turbulence, is low.
While both models can achieve the high accretion rates required for massive star formation, the specific initial conditions of the gas out of which they form will set the stage for the fragmentation properties of pre-stellar cores and the accretion history of massive stars. 

One way to determine the initial conditions of massive star formation is to study the demographics of massive starless clumps and cores. \add{Most studies have found that massive cores are typically subvirial and should collapse within a gravitational free-fall time if the cores are not supported by magnetic fields} \citep[\add{e.g.,}][]{Motte2007a, Pillai2011a, Kauffmann2013a, Sanchez-Monge2013a, Battersby2014a, Lu2015a, Zhang2015a, Henshaw2016a, Contreras2018a, LiuLi2018a, Traficante2018a, Traficante2018b, Williams2018a}. \add{In contrast, there have only been a few observational studies that have discovered virialized candidates \citep[e.g.,][]{Tan2013a, Kainulainen2013a, LiuLi2018a}}. If most massive cores are subvirial, then magnetic fields may provide additional support against collapse. Observations suggest that strong magnetic fields of the order of $\sim1$ mG are required for stabilizing massive pre-stellar cores. However, observations have demonstrated that magnetic pressure in dense, molecular gas is dynamically sub-dominant to gravity \citep{Crutcher2010a}. 



The dynamic state of massive cores will affect how mass is gathered to the star. Is most of the mass gathered via the direct global collapse of subvirial cores or is it slowly accumulated from the turbulent collapse of roughly virialized cores? The purpose of this work is to understand how the initial state of the pre-stellar core affects the growth of the resulting massive star and fragmentation of the core. However, massive stars are rare and form in highly embedded regions, therefore capturing the early moments of their formation and following how their mass is accreted onto the star is observationally challenging. They also have large luminosities throughout their formation since they contract to the main sequence while they are actively accreting and therefore radiation pressure is an important feedback mechanism during their formation \citep{Larson1971a, Wolfire1987a}. Instead, we must turn to multi-dimensional radiation-hydrodynamic (RHD) numerical simulations to study the early formation of massive stars. 

\add{Additionally, the accretion rate onto the massive star throughout the star formation process and the degree of core fragmentation into companions is likely dependent on the initial core properties such as the initial degree of turbulence and virial parameter.  We investigate this effect here by performing three-dimensional RHD numerical simulations of the collapse of subvirial and virialized turbulent massive pre-stellar cores. The simulations presented in this work are still highly idealized since we do not include magnetic fields or outflows, which will be addressed in future work. This paper is organized as follows: we describe our numerical methodology and simulation design in Section~\ref{sec:methods}, we present and discuss our results in Sections~\ref{sec:res} and \ref{sec:disc}, respectively. Finally, we conclude in Section~\ref{sec:conc}.}

\section{Simulation Details}
\label{sec:methods}
In this paper, we simulate the gravitational collapse of isolated turbulent massive pre-stellar cores using the \orion\ adaptive mesh refinement (AMR) code \citep{Li2012a} to understand how their stability affects their fragmentation and the formation history of massive stars. We summarize the numerical methods, including the equations we solve in the \orion\ code, our refinement criteria, and how we treat the stellar radiation field in Section~\ref{sec:numeth}. We present the initial and boundary conditions for our simulations in Section~\ref{sec:ics}. We refer the reader to \pap\ for more details on the numerics and overall algorithm (e.g., see Section~2.5 of \pap) since the physics included and the numerical methods used in this work are identical except where specified below. The initial conditions and numerical parameters are summarized in Table~\ref{tab:sim}.

\subsection{Numerical Methods and Refinement Criteria}
\label{sec:numeth}
 \orion\ uses the \textsc{Chombo} toolset to solve partial differential equations on block-structured AMR meshes \citep{Chombo2015a} and solves the equations of gravito-radiation-hydrodynamics in the two-temperature, mixed-frame flux-limited diffusion (FLD) approximation on a Cartesian adaptive grid \citep{Krumholz2007a}. These equations are given by 
\begin{equation}
\label{eqn:com}
\frac{\partial \rho}{\partial t} =  -\mathbf{\nabla} \cdot \left( \rho \mathbf{v} \right) - \sum_{i}  \dot{M}_i W(\mathbf{x} - \mathbf{x}_i)
\end{equation}
\begin{eqnarray}
\frac{\partial \left(\rho \mathbf{v} \right)}{\partial t}  &=& -\mathbf{\nabla} \cdot \left( \rho \mathbf{v}  \bf{v} \right) - \mathbf{\nabla} P - \rho \mathbf{\nabla} \phi - \lambda \mathbf{\nabla} E_{\rm R} \nonumber \\
	&&+ \sum_i \left[\dot{\mathbf{p}}_{\rm rad, \it i}-\dot{\mathbf{p}}_{\it i}W(\mathbf{x} - \mathbf{x}_i)\right]
\label{eqn:cop}
\end{eqnarray}
\begin{eqnarray}
\frac{\partial \left( \rho e\right)}{\partial t} &=& - \mathbf{\nabla} \cdot \left[ (\rho e + P)\textbf{v} \right] - \rho \mathbf{v} \cdot \mathbf{\nabla} \phi - \kappa_{\rm 0P} \rho(4\pi B - cE_{\rm R} ) \nonumber \\ 
	&&+ \lambda \left( 2 \frac{\kappa_{\rm 0P}}{\kappa_{\rm 0R}} - 1 \right) \mathbf{v} \cdot \mathbf{\nabla} E_{\rm R} - \left(\frac{\rho}{m_{\rm p}} \right)^2 \Lambda(T_{\rm g}) \nonumber \\
	&&+ \sum_i \left[\dot{\mathbf{\varepsilon}}_{\rm rad, \it i}-\dot{\varepsilon}_i W(\mathbf{x} - \mathbf{x}_i) \right]
\label{eqn:coe}
\end{eqnarray}
\begin{eqnarray}
\frac{\partial E_{\rm R}}{\partial t} &=& \mathbf{\nabla} \cdot \left( \frac{c \lambda}{\kappa_{\rm 0R} \rho} \mathbf{\nabla} E_{\rm R} \right) + \kappa_{\rm 0P} \rho \left(4\pi B - c E_{\rm R} \right) \nonumber \\
	&&- \lambda \left( 2 \frac{\kappa_{0 \rm P}}{\kappa_{0 R}} - 1\right) \mathbf{v} \cdot \nabla E_{\rm R} 
	- \nabla \cdot \left( \frac{3 - R_2}{2} \mathbf{v} E_{\rm R}\right) \nonumber \\ 
	& & +  \left(\frac{\rho}{m_{\rm p}} \right)^2 \Lambda(T_{\rm g}). 
\label{eqn:coer}
\end{eqnarray}

\begin{equation}
\label{eqn:mdot}
\frac{dM_{\rm i}}{dt} = \dot{M}_{\rm i}
\end{equation}

\begin{equation}
\label{eqn:dvi}
\frac{d \bf{x}_{i}}{dt} = \frac{\bf{p}_{\rm i}}{M_{\rm i}}
\end{equation}

\begin{equation}
\label{eqn:dpi}
\frac{d\bf{p}_i}{dt} = -M_i \nabla \phi + \dot{\bf{p}}_i
\end{equation}

\begin{equation}
\label{eqn:pois}
\nabla^2 \phi = 4 \pi G \left[ \rho + \sum_i M_i \delta(\bf{x} - \bf{x}_i)\right].
\end{equation}
\noindent
Here, $\rho$ is the gas density,  $\rho \bf{v}$ is the gas momentum, $\rho e$ is the total internal plus kinetic gas energy density, and $E_{\rm R}$ is the radiation energy density in the rest frame of the computational domain, $\kappa_{\rm 0P}$ and $\kappa_{\rm 0R}$ are the Planck and Rosseland mean opacities of the dust-plus-gas fluid, $B$ is the blackbody function,  $\Lambda$ is the rate of gas cooling that takes into account line and continuum processes for gas at temperatures above $\gtrsim 10^3$ K, $\lambda$ is the flux-limiter, and $R_2$ is the Eddington factor. For more information on the flux-limiter, Eddington factor, hot gas cooling rate, and choice of dust opacities, we refer the reader to \pap. 

In addition to the fluid, \orion\ contains Lagrangian sink particles used to represent radiating (proto)stars, indexed by subscript $i$, where each particle $i$ is described with a mass $m_{\rm i}$, position $\bf{x}_{i}$, and momenta $\bf{p}_{i}$. Additionally, the particles accrete gas at a rate $\dot{M}_{\rm i}$ within four fine-level cells (i.e., the accretion around each star particle as weighted by the weighting kernel $W(\mathbf{x}-\mathbf{x}_i)$ described in \citet{Krumholz2004a}).  We refer the reader to \pap\ for our description of our merging criterion for star particles when they pass within one accretion radius of each other. 

Equations~\ref{eqn:com}-\ref{eqn:coer} describe conservation of gas mass, gas momentum, gas total energy, and radiation total energy. Equations~\ref{eqn:mdot}-\ref{eqn:dpi} describe how the star particles mass, velocity, and momenta are updated as they accrete mass and interact gravitationally with the surrounding fluid. Here $\phi$ is the gravitational potential that obeys the Poisson equation, given by Equation~\ref{eqn:pois}, that includes contributions from both the fluid and star particles. We also assume an ideal equation of state so that the gas pressure is 
\begin{equation}
P=\frac{\rho k_{\rm B}T}{\mu m_{\rm H}} = \left( \gamma-1\right) \rho e_{\rm T},
\end{equation}
where $T$ is the gas temperature, $\mu$ is the mean molecular weight,  $\gamma$ is the ratio of specific heats, and $e_{\rm T}$ is the thermal energy of the gas per unit mass. We take $\mu=2.33$  and $\gamma=5/3$ that is appropriate for molecular gas of solar composition at temperatures too low to excite the rotational levels of H$_2$. The fluid is a mixture of gas and dust, and at the high densities that we are concerned with the dust is thermally coupled to the gas, allowing us to assume that the dust temperature is the same as the gas temperature.

\begin{table*}
	\begin{center}
	\caption{
	\label{tab:sim}
Simulation Parameters
}
	\begin{tabular}{ l c  c  c  c  c  c  c c}
	\hline
	\textbf{Run} & & \subvir\ & \vir\ \\
	\\
	\hline
	\textbf{Initial Physical Parameters}\\
	\hline
	\hline
	\textbf{Virial Parameter} \tablenotemark{a} & $\alpha_{\rm vir}$ & 0.14 & 1.1\\
	Cloud Mass [$\rm M_{\rm \odot}$] & $M_{\rm c}$ & 150 & 150 \\
	Cloud Radius [pc] & $R_{\rm c}$ & 0.1 & 0.1 \\
	Surface Density [$\rm g \, cm^{-2}$] & $\Sigma$ & 1 & 1\\
	EOS Index \tablenotemark{b} & $n$ & 5/3  & 5/3 \\
	Temperature [K] & $T_{\rm c}$ & 20 & 20\\
	Sound speed [$\rm km \, s^{-1} $] & $c_s$ & 0.27 & 0.27 \\
	Mean Density [$\rm 10^{-18}\, g \, cm^{-3}$] & $\bar{\rm \rho}_{\rm cl}$ & $2.4$ & $2.4$ \\
	Mean Free-fall Time  [kyr] & $t_{\rm ff}$ & 42.8 & 42.8\\
	Power Law Index & $\kappa_{\rm \rho}$ & 1.5 & 1.5 \\
	3D Velocity Dispersion [$\rm km \, s^{-1} $] \tablenotemark{c}& $\langle \sigma_{\rm 3D}\rangle_V$ & 0.73 & 2.1\\
	1D Velocity Dispersion [$\rm km \, s^{-1} $] \tablenotemark{c, \textrm d} & $\langle \sigma_{\rm 1D}\rangle_V$ & 0.42 & 1.2\\
	Mach Number\tablenotemark{e} & $\mathcal{M}$ & 1.6 & 4.5 \\
	Specific Angular Momentum [$\rm \rm 10^{19} \, cm^2 \, s^{-1}$] & $J_{\rm spec}$ & 5.00  & 21.6 \\
	\hline
	\\
	\textbf{Numerical Parameters}\\
	\hline
	\hline
	Domain Length [pc] & $L_{\rm box}$ & 0.4 & 0.4 \\
	Base Grid Cells & $N_{\rm 0}$ & $128^3$ & $128^3$ \\ 
	Maximum Level & $l_{\rm max}$ & 5  & 5\\
	Minimum Cell Size [AU] & $\Delta x_{\rm l_{\rm max}}$ &  20 & 20  \\
	Jeans Length Refinement &$J_{\rm max}$ & 0.125 & 0.125 \\
	$E_{\rm R}$ Gradient Refinement & $E_{\rm R}/\Delta x$ & 0.15 & 0.15  \\
	Accretion Radius [AU] &  $4 \Delta x_{\rm l_{\rm max}}$ & 80 & 80\\ 
	\hline
	\\
	\textbf{Simulation Outcomes} \\
	\hline
	\hline
	Simulation Time [$t_{\rm ff}$] & & 0.87 & 0.96  \\
	Massive Star Mass [$\rm M_{\odot}$] & & 61.7 & 51.95 \\
	Number of Sinks \tablenotemark{f} && 3 & 18 \\
	Total Companion Star Mass [$\rm M_{\odot}$] && 0.25 & 14.1 \\
	\end{tabular}
	\tablenotetext{a}{ $\alpha_{\rm vir} = 5 \langle \sigma_{\rm 1D}\rangle_V^2 R_{\rm c}/G M_{\rm c}$}
	\tablenotetext{b}{ Equation of state: $P \propto \rho^{n}$.} 
	\tablenotetext{c}{Volume-weighted}
	\tablenotetext{d}{ $\sigma_{\rm 1D} = \sigma_{\rm 3D}/\sqrt{3}$} 
	\tablenotetext{e}{ $\mathcal{M} = \langle \sigma_{\rm 1D} \rangle_V/c_s$} 
	\tablenotetext{f}{Final number of sinks with masses greater than 0.01 $M_{\rm \odot}$.}
	\end{center}
\end{table*}

Each star particle includes a model for protostellar evolution used to represent them as radiating protostars with frequency dependent stellar luminosities \citep{Lejeune1997a, Hosokawa2009a, Offner2009a, Rosen2016a}. In order to properly model both the direct and indirect radiation pressure we use the \harm\ algorithm presented in \citet{Rosen2017a}. \harm\ uses a frequency-dependent adaptive long-characteristics ray-tracing scheme that accurately solves the radiative transfer equation along rays to model the radial stellar radiation field from stars that is absorbed by dust and is coupled to a grey moment method, in our case FLD, which models the dust re-processed thermal radiation field intrinsic to the dusty fluid. The energy and momentum absorbed from the stellar radiation field, including contributions from the accretion luminosity,
\begin{equation}
L_{\rm acc} = f_{\rm rad} \frac{ G M_{\rm \star} \dot{M}_{\rm \star}}{R_{\rm \star}},
\end{equation}
by the dust are added at a rate of $\dot{\epsilon}_{\rm{rad}, i}$ and $\dot{\bf{p}}_{\rm{rad}, i}$ for particle $i$ as shown in equations~\ref{eqn:cop}-\ref{eqn:coe}. Here $f_{\rm rad}$ is the fraction of the gravitational potential energy of the accretion flow that is converted to radiation and we take $f_{\rm rad}=3/4$ following \citet{Offner2009a} and $M_{\rm \star}$ and $R_{\star}$ are the star's mass and radius, respectively. We use the frequency dependent dust opacities from \citet{Weingartner2001a} and divide the stellar spectrum and dust opacities into ten frequency bins (e.g., see Figure~1 in \pap). To reduce the computational cost, the rays adaptively split to conserve solid angle as they propagate radially away from their sources. We refer the reader to \citet{Krumholz2007a}, \pap, and \citet{Rosen2017a} for a complete description of our treatment of the direct and indirect radiation pressures modeled in this work.

For each simulation we begin with a base grid with volume (0.4 pc)$^3$ discretized by $128^3$ cells and allow for five levels of refinement resulting in a maximum resolution of 20 AU.  As the simulation evolves, the AMR algorithm automatically adds and removes finer grids based on certain refinement criteria set by the user. Following \pap, we refine cells if they meet at least one of the following criteria: (1) any cell on the base level (i.e., level 0) that has a density equal to or greater than the core's edge density so that the entire prestellar core is refined to level 1; (2) any cell where the density in the cell exceeds the Jeans density given by 
\begin{equation}
\label{eqn:rhoj}
\rho_{\rm max,J } = \frac{\pi J^2_{\rm max} c_{\rm s}^2}{G \Delta x_l^2}
\end{equation}
where $c_{\rm s}= \sqrt{k_B T/(\mu m_{\rm H})}$ is the isothermal sound speed, $\Delta x_l$ is the cell size on level $l$, and $J_{\rm max}$ is the maximum allowed number of Jeans lengths per cell, which we set to 1/8  \citep{Truelove1997a}; (3) any cell that is located within at least 8 cells of a sink particle; and (4) any cell within which the radiation energy density gradient exceeds $\left|\nabla E_{\rm R}\right| > 0.15 E_{\rm R}/\Delta x_{\rm l}$ (i.e., where the radiation energy density changes by more than 15\% over the size of a single cell). This procedure is repeated recursively on all levels after every two level updates.  Finally, a sink particle can only be created when the Jeans density is violated on the finest level in which we take $J_{\rm max}=1/4$ in Equation~(\ref{eqn:rhoj}) following the artificial fragmentation tests of \citet{Truelove1997a} \addd{that found that values of $J\lesssim 1/4$ causes artificial fragmentation to be avoided in gravitational collapse simulations.}
When a cell on the finest level exceeds the Jeans density we place a sink particle in that cell whose mass is the excess matter in that cell. The new sink particle will evolve according to equations~\ref{eqn:mdot}-\ref{eqn:dpi}.

\subsection{Initial and Boundary Conditions}
\label{sec:ics}
In this work, we perform two simulations that are identical in every way except for their initial one-dimensional velocity dispersion ($\sigma_{1D}$) in order to determine how the dynamic state of the star-forming core (i.e., virialized versus subvirial) affects their fragmentation and the growth of massive stars. Following \pap, we begin with an isolated pre-stellar core of molecular gas and dust,  where we assume a dust-to-gas ratio of 0.01, with mass $M_{\rm c}=150 \; M_{\rm \odot}$, radius $R_{\rm c} = 0.1$ pc, and initial gas temperature of 20 K corresponding to a surface density of $\Sigma = M_{\rm c}/\pi R^2_{\rm c} = 1 \; \rm{g \; cm^{-2}}$ consistent with \add{extreme} massive star-forming environments \add{such as W49, Sgr B2, W51, and Cygnus X1} and massive core densities \citep[\add{e.g.,}][]{Galvan2013a, Ginsburg2015a, Ginsburg2018a, Cao2019a}. The corresponding mean density of the core is $\bar{\rho} = 2.4 \times 10^{-18} \; \rm{g \; cm^{-3}}$ ($1.2 \times 10^6 \; \rm{H \; nuclei \; cm^{-3}}$) and its characteristic free-fall collapse time scale is $t_{\rm ff}\approx 42.6 \; \rm{kyr}$.  The core follows a density profile $\rho(r) \propto r^{- 3/2}$ in agreement with observations of massive cores at the $\sim$0.1 pc scale and clumps at the $\sim$1 pc scale that find values of $\kappa_{\rm \rho} = 1.5-2$ \citep[e.g.,][]{Caselli1995a, Beuther2002b, Mueller2002a, Beuther2007a, Zhang2009a, Longmore2011a, Butler2012a, Battersby2014a, Stutz2016a}. Each core is placed in the center of a 0.4 pc box that is filled with hot, diffuse gas with density $\rho_{\rm amb} = 0.01 \rho_{\rm edge}$ where $\rho_{\rm edge}$ is the density at the core boundary and temperature $T_{\rm amb} = 2000$ K so that the core \add{is isolated and} is in \add{thermal} pressure balance with the ambient medium. 
We set the opacity of the ambient medium to zero \add{(i.e., the ambient medium is dust-free)}. 
\begin{figure*}
\centerline{\includegraphics[trim=0.2cm 0.2cm 0.2cm 0.2
cm,clip,width=0.85\textwidth]{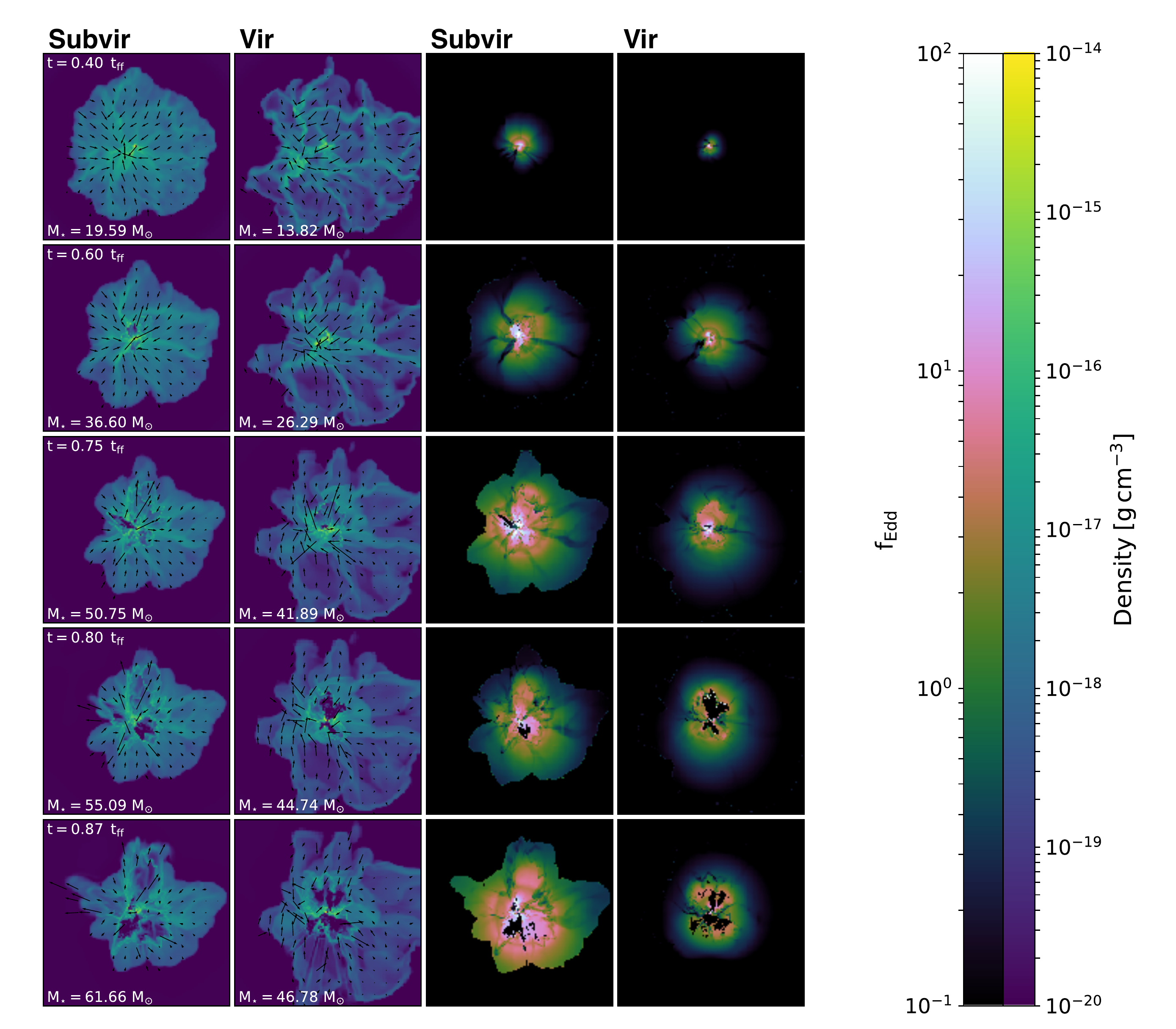}}
\caption{
\label{fig:fedd}
Density slices along the $yz$-plane of a collapsing, turbulent pre-stellar core into a massive stellar system for runs \subvir\ (far left column) and \vir\ (center left column) with velocity vectors over plotted. The velocity vectors are scaled as $\sqrt{v}$ in units of km/s and we only over plot vectors for densities $ \ge 5 \times 10^{-20} \; \rm{g \; cm^{-3}}$. The two right columns show the corresponding slices of the Eddington ratios ($f_{\rm Edd} = F_{\rm rad}/F_{\rm grav}$) for runs \subvir\ (middle right column) and \vir\ (right-most column). Each panel is (50 kAU)$^2$ with the center of each panel corresponding to the location of the most massive star that has formed.  The time of the simulation and mass of the most massive star are given in the top-left corner of the far left panels and the bottom left corner of the two left panels of each row, respectively. 
}
\end{figure*}

We include turbulence by seeding the initial gas velocities ($v_x$, $v_y$, and $v_z$) with a velocity power spectrum, $P(k) \propto k^{-2}$, as is expected for supersonic turbulence \citep{Padoan1999a, Boldyrev2002a, Cho2003a, Kowal2007a}. We include modes between $k_{\rm min} = 1$ to $k_{\rm max} = 256$ and take the turbulence mixture of gas to be 1/3 compressive and 2/3 solenoidal, which is consistent for the natural mixture of a 3D fluid \citep{Kowal2007a, Kowal2010a}. The onset of turbulence modifies the density distribution and we allow the turbulence to decay freely. For both simulations, we use the same velocity perturbation power spectrum at initialization but set the initial $\sigma_{\rm 1D}$ amplitude to two different values: our virial run, \vir, has $\sigma_{\rm 1D}= 1.2$ km/s corresponding to $\alpha_{\rm vir} = 1.1$ and $\mathcal{M} = \langle \sigma_{\rm 1D} \rangle_m/c_s = 4.4$ and our subvirial run, \subvir, is initialized with $\sigma_{\rm 1D}= 0.42$ km/s yielding $\alpha_{\rm vir} = 0.14$ and $\mathcal{M} =1.6$.\footnote{We note that run \subvir\ is the same run as \texttt{TurbRT+FLD} from \pap.} \add{We note that allowing the turbulence to decay is somewhat unrealistic. However, this simplification should have little effect on our results since the decay timescale, $\approx D/\sigma_{\rm 1D}$ where $D$ is the core diameter \citep{Goldreich1995a}, is $\sim$0.16 Myr and $\sim$0.46 Myr for runs \vir\ and \subvir, respectively; which are much longer than the runtime for both simulations presented in this work.} \addd{Therefore, we expect that allowing the turbulence to decay should have a negligible effect on our results.} 

Our boundary conditions for the radiation, gravity, and hydrodynamic solvers are as follows. For each radiation update, we impose Marshak boundary conditions that bathe the simulation volume with a blackbody radiation field equal to $E_0 = 1.21 \times 10^{-9} \rm{\; erg \; cm^{-3}}$ corresponding to a 20 K blackbody but we allow radiation generated within the simulation volume to escape freely \citep{Krumholz2009a, Cunningham2011a, Myers2013a, Rosen2016a}. We set the gravitational potential, $\phi$, to zero at all boundaries when solving equation~\ref{eqn:pois} \citep{Myers2013a}. Since the core boundaries are far removed from the domain boundaries we do not expect this choice of boundary conditions to lead to any significant square artifacts near the domain boundaries. Finally, we impose outflow boundary conditions for the hydrodynamic update by setting the gradients of the hydrodynamic quantities $\left(\rho, \; \rho \bf{v}, \rho e \right)$ to be zero at the domain when advancing equations~\ref{eqn:com}-\ref{eqn:coe} \citep{Cunningham2011a, Myers2013a}.

\begin{figure*}
\centerline{\includegraphics[trim=0.2cm 0.2cm 0.2cm 0.2cm,clip,width=1\textwidth]{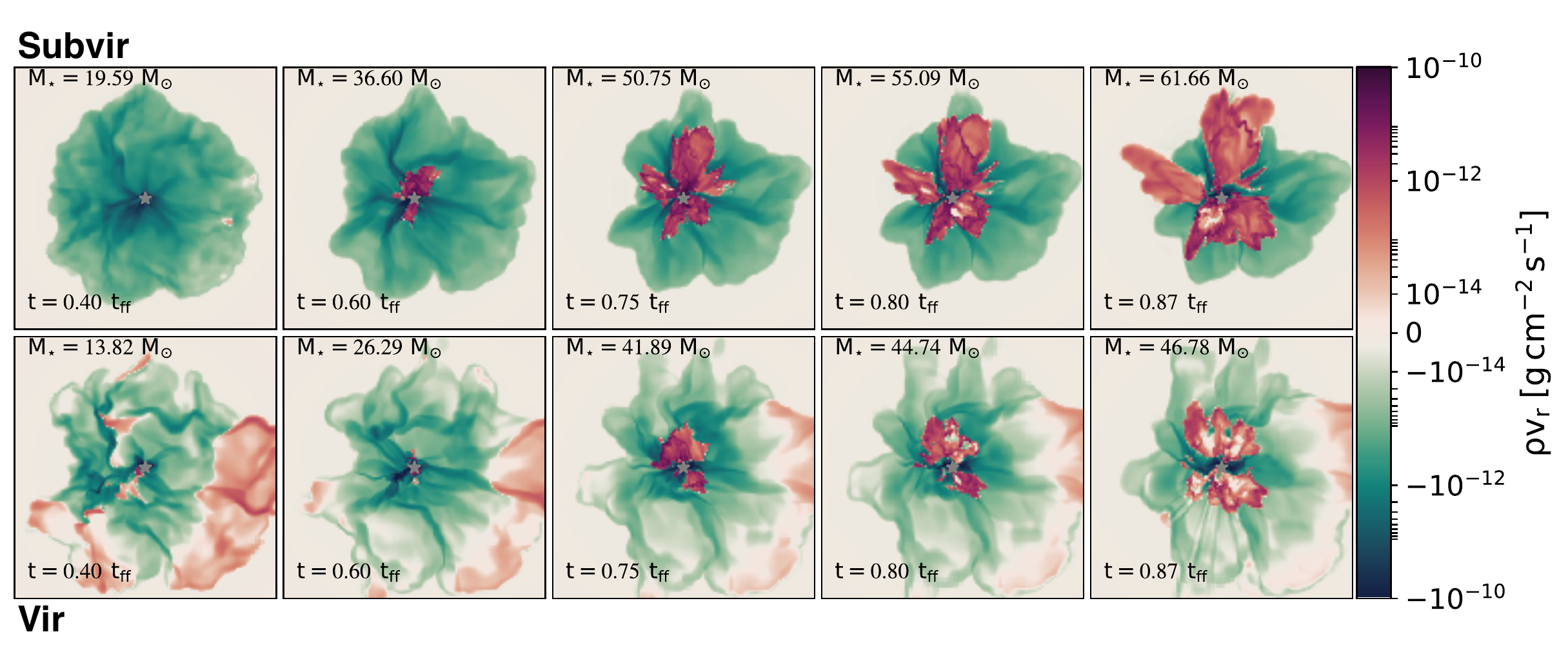}}
\caption{
\label{fig:rhovr}
Thin projections of the gas radial velocity with respect to the primary star multiplied by the gas density to show material that is moving towards (negative values of $\rho v_r$) and away from (positive values of $\rho v_r$)  the star for runs \subvir\ (top panels) and \vir\ (bottom panels) at different times. Each panel is (50,000 AU)$^2$ in area and the projection is taken over a depth of 1000 AU in front of and behind the massive star. The gray star at the center of each panel denotes the location of the massive star.
}
\end{figure*}

\section{Results}
\label{sec:res}
In this section, we describe and compare our results for runs \subvir\ and \vir\ described in Section~\ref{sec:ics} and summarized in Table~\ref{tab:sim}. Run \subvir\ was run on the Hyades supercomputer located at UC Santa Cruz and run \vir\ was run on the NASA supercomputer Pleiades located at NASA Ames. \add{We run each simulation to the point where the simulation takes too long to evolve because the majority of the bubble shells are refined to the finest level, severely increasing the computational cost of the simulation.} We use the \textsc{yt} package \citep{Turk2011a} to produce all the figures and quantitative analysis shown below.

\subsection{Core Collapse Properties and  Influence of Radiative Feedback}
\label{sec:collapse}

We show a series of the density slices and the corresponding Eddington ratio slices, $f_{\rm Edd}= F_{\rm rad}/F_{\rm grav}$ where $F_{\rm rad}$ is the total radiation force from both the stellar and dust-reprocessed radiation fields and $F_{\rm grav}$ is the gravitational force, for runs \subvir\ and \vir\ in Figure~\ref{fig:fedd}. The two left hand columns of Figure~\ref{fig:fedd} show the density structure and collapse evolution along the $yz-$plane with velocity vectors over-plotted for runs~\subvir\ (left-most column) and \vir\ (center left column), respectively. These panels show that the pre-stellar core in run~\subvir\ undergoes a rapid, monolithic collapse whereas the core in run~\vir\ undergoes a gradual collapse leading to a slower growth rate for the massive star. At the end of run \subvir, $t=0.87 \; t_{\rm ff}$, the primary star has a final mass of 61.66 $M_{\rm \odot}$ whereas for run \vir\ the primary star is significantly lower in mass with a mass of 46.78 $M_{\rm \odot}$ at this time. This result \add{suggests} that the collapse of subvirial cores\add{, as compared to virialized cores, may} lead to massive stars that have larger masses at birth\add{.}

The slower growth of the primary star in run \vir\ occurs because the core envelope has a larger velocity dispersion and angular momentum content resulting in greater support against direct gravitational collapse. This is seen in the velocity vectors over-plotted on the density slice plots that show a randomized orientation in the core envelope whereas in run \subvir\ the velocity vectors point radially towards the massive star demonstrating that the core collapses monolithically. Instead, for run \vir, the velocity vectors in the core envelope\add{'s} low density regions are oriented towards the high density flattened filaments that form out of the core's turbulence at early times\add{. This effect} caus\add{es} the envelope material to be accreted onto filaments rather than fall directly towards the primary star.

\begin{figure*}
\centerline{\includegraphics[trim=0.0cm 0.0cm 0.0cm 0.0cm,clip,width=0.9\textwidth]{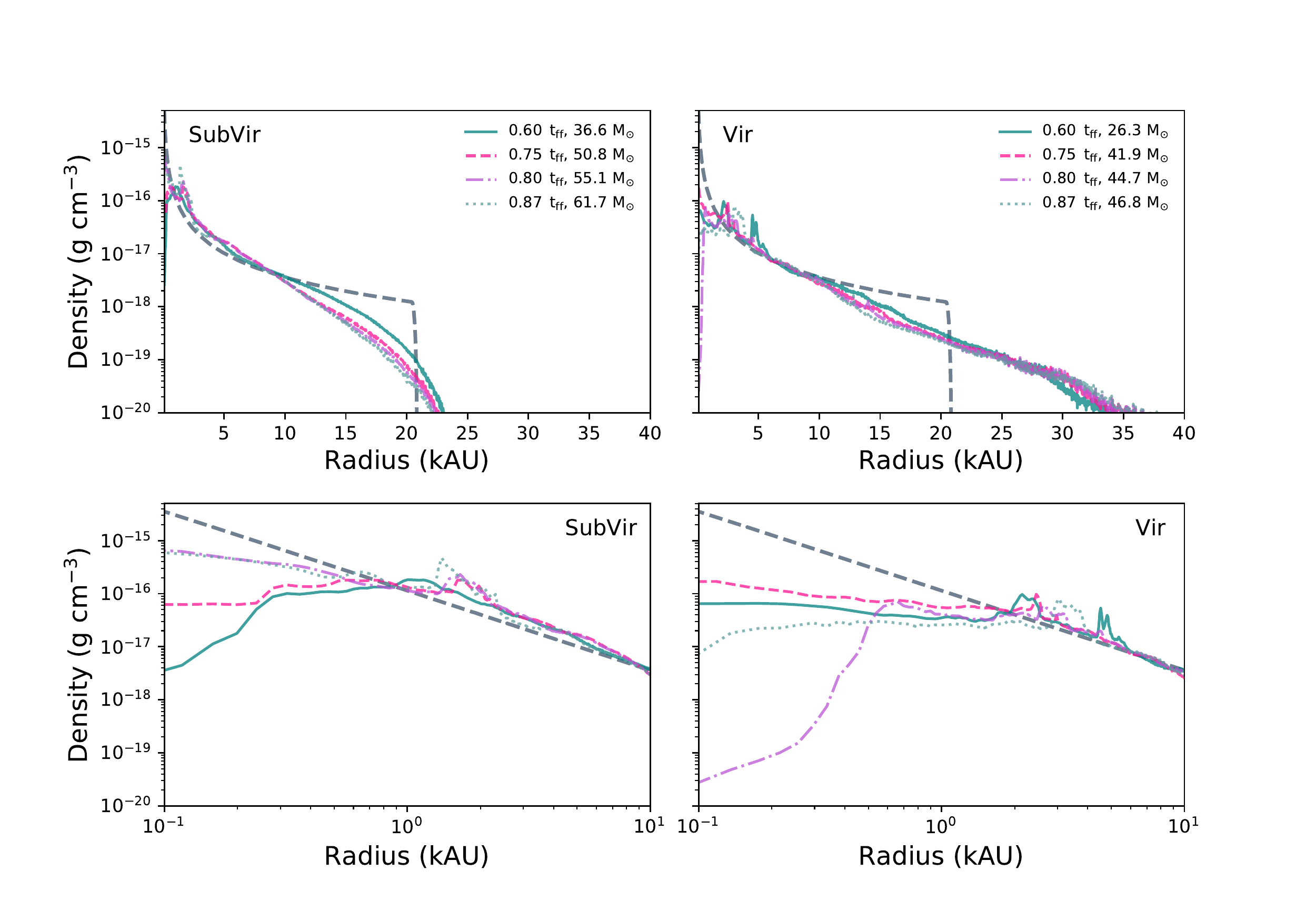}}
\caption{
\label{fig:rho_pro}
Density profiles for runs \subvir\ (left panels) and \vir\ (right panels). The dashed gray line in both panels shows the initial density distribution of the core and the legend gives the mass of the primary star and simulation time. The top panels show these profiles on a linear-log scale to highlight the density distribution of the core envelope. The bottom panels are plotted in log-log scale to highlight the density structure of the inner part of the core.
}
\end{figure*}

At high stellar masses the radiation pressure associated with the massive star's high luminosity launches radiation pressure dominated bubbles that expand away from the star, reducing the infall of material to the accreting primary star \citep{Yorke2002a, Krumholz2009a, Kuiper2011a, Rosen2016a}. When this occurs an optically thick accretion disk has formed around the massive star due to conservation of angular momentum as the core collapses and the bubbles are launched above and below the accretion disk as material is funnelled to the massive star. These low-density expanding bubbles are seen in the density slice plots for both runs in Figure~\ref{fig:fedd} at late times. We find that they are not sustained until the primary stellar mass reaches $\gtrsim 30 M_{\rm \odot}$ for both runs because the ram pressure from the collapsing core material quenches the bubbles at early times. However, as the stellar mass of the primary star, and therefore its luminosity, increases these bubbles expand and grow larger with time. These bubbles occur in regions where $f_{\rm Edd} \gg 1$  as shown in the two right-hand columns of Figure~\ref{fig:fedd}. These panels show that at early times ($t=0.4 \; t_{\rm ff}$), when the star is 19.59 $M_{\rm \odot}$ (13.82 $M_{\rm \odot}$) for run \subvir\ (\vir) the inner regions of the core nearest the primary star are super-Eddington (i.e., $f_{\rm Edd}>1$) but that the radiation pressure dominated bubbles have not been launched yet for either run at this point. This is because the acceleration of the material is initially controlled by the gravitational collapse of the core material and is falling towards the star. Thus, it takes time for the radiation pressure to overturn the accretion flow onto the star even though the gas near the star is super-Eddington. 

As the primary star continues to grow in mass the region of the core that becomes super-Eddington increases in size. This is more apparent for run \subvir, which leads to the majority of the core to become super-Eddington at late times because the star is very massive and therefore extremely luminous with a zero-age main sequence luminosity of $L_{\rm \star} = 5.38 \times 10^{5} \; L_{\rm \odot}$ \citep{Tout1996a}. The $f_{\rm Edd}$ panels shown in Figure~\ref{fig:fedd} also show that the dense filaments that are inherent to the turbulent structure of the core for both runs achieve $f_{\rm Edd} \lesssim 1$. Thus, these filaments are able to self-shield from the strong radiation pressure associated with the primary star and can be incorporated into the star-disk system as the core collapses. 

To further illustrate the gravitational collapse of the core and the launching of the low-density bubbles driven by radiative feedback from the primary star we show a series of thin projection plots of the radial momentum, $\rho v_{\rm r}$, of gas with respect to the primary star for runs~\subvir\ (top row) and \vir\ (bottom row) in the $yz-$plane in Figure~\ref{fig:rhovr}. Each projection is taken over a depth of 1000 AU with the primary star at the center of each panel. Values of positive $\rho v_{\rm r}$ denote gas that is moving away from the primary star whereas negative values denote gas that is falling towards it. These snapshots show the growth and expansion of the radiation pressure dominated bubbles is perpendicular to the accretion disk. The accretion disk is larger in size in run \vir\ since the pre-stellar core has a larger angular momentum content due to its larger initial velocity dispersion (e.g., see Table~\ref{tab:sim}). This `flashlight' effect allows material to be funnelled to the star by the accretion disk, as demonstrated by the negative values of $\rho v_{\rm r}$ along the equatorial plane of the star, while the radiative flux escapes along directions above and below the disk \citep{Yorke2002a}. We describe the accretion disk structure and evolution in more detail in Section~\ref{sec:acc}.

For run \subvir, Figure~\ref{fig:rhovr} shows that the core undergoes a more significant collapse than run \vir\ in agreement with the density slice plots in Figure~\ref{fig:fedd} and that the full core, at early times, is collapsing overall.  This leads to an overall higher density for the core in run \subvir\ than the core in run \vir\ as the simulations progress. In contrast, the radial momentum plots for run \vir\ show that while the majority of the core, especially the inner parts, are collapsing towards the star a significant portion of the outer envelope is expanding. This is due to the higher degree of turbulence providing support against collapse.  As the envelope expands, its velocity and density decreases with time. Eventually the majority of the envelope begins to collapse towards the primary star at late times. This Figure also shows that the turbulence produces dense, optically thick filaments as the core collapses. Since the core in run \subvir\ has a low velocity dispersion and collapses immediately the filaments are typically wider than those that are produced in run \vir. 

The evolution of the core expansion is shown in Figure~\ref{fig:rho_pro}, which shows the spherically averaged density profiles of the core as a function of radius  for run \subvir\ (top left panel) and run \vir\ (top right panel). The gray dashed line in each panel denotes the initial density distribution of the core before the turbulent velocity structure is added to the gas velocities (i.e., at $t=0$), described in Section~\ref{sec:ics}. These panels present the density in $\log$ scale and the radius in linear scale to show the overall density evolution of the cores and how turbulence can modify the density distribution. As these panels demonstrate the onset of turbulence causes the cores to expand and the core gas at the core boundary mixes with the ambient medium. While this effect is a limitation due to treating our core as an isolated object in pressure balance with a low-density ambient medium it should have little effect on our results since the cores undergo inside-out collapse \citep{Shu1977a}. For run~\subvir\ the expansion due to the onset of turbulence is small, only increasing by $\sim$3 kAU (15\% increase) in size. However, the onset of turbulence for run \vir\ causes the overall core to expand to $\sim$30 kAU (50\% increase) at early times and continues to expand as the simulation progresses. This is because the initial global velocity dispersion of the core in run~\vir\ is a factor of $\sim3$ larger than run \subvir. As these panels demonstrate, the core in run \vir\ undergoes a less rapid inside-out collapse than run \subvir\ because of the greater support from turbulence. This causes the inner core collapse in run \vir\ to be more gradual leading to a slower growth rate for the primary star. 

The bottom panels in Figure~\ref{fig:rho_pro} show a zoom-in of the inner density distribution of the cores in run \subvir\ (bottom left panel) and run \vir\ (bottom right panel) but now both panels are in log-log scale to highlight the density distribution near the primary star.  The inner regions of the core collapse for both runs and this gas is incorporated into the massive star or is blown away due to the strong radiation pressure from the massive star. We find that the inner density profile of the core for run \vir\ is significantly larger in density at $t=0.60 \; t_{\rm ff}$ as compared to run \subvir\ for distances $\lesssim 300$ AU from the massive star. This is because the radiation pressure dominated bubbles have not been launched yet at this time for run \vir\ but they have been launched in run \subvir\ due to the larger stellar mass attained in a shorter amount of time (e.g., $M_{\rm \star}=36.6 \; M_{\rm \odot}$ for run \subvir\ as compared to $M_{\rm \star}=26.3 \; M_{\rm \odot}$ for run \vir). 

\begin{figure}
\centerline{\includegraphics[trim=0.0cm 0.0cm 0.0cm 0.0cm,clip,width=0.95\columnwidth]{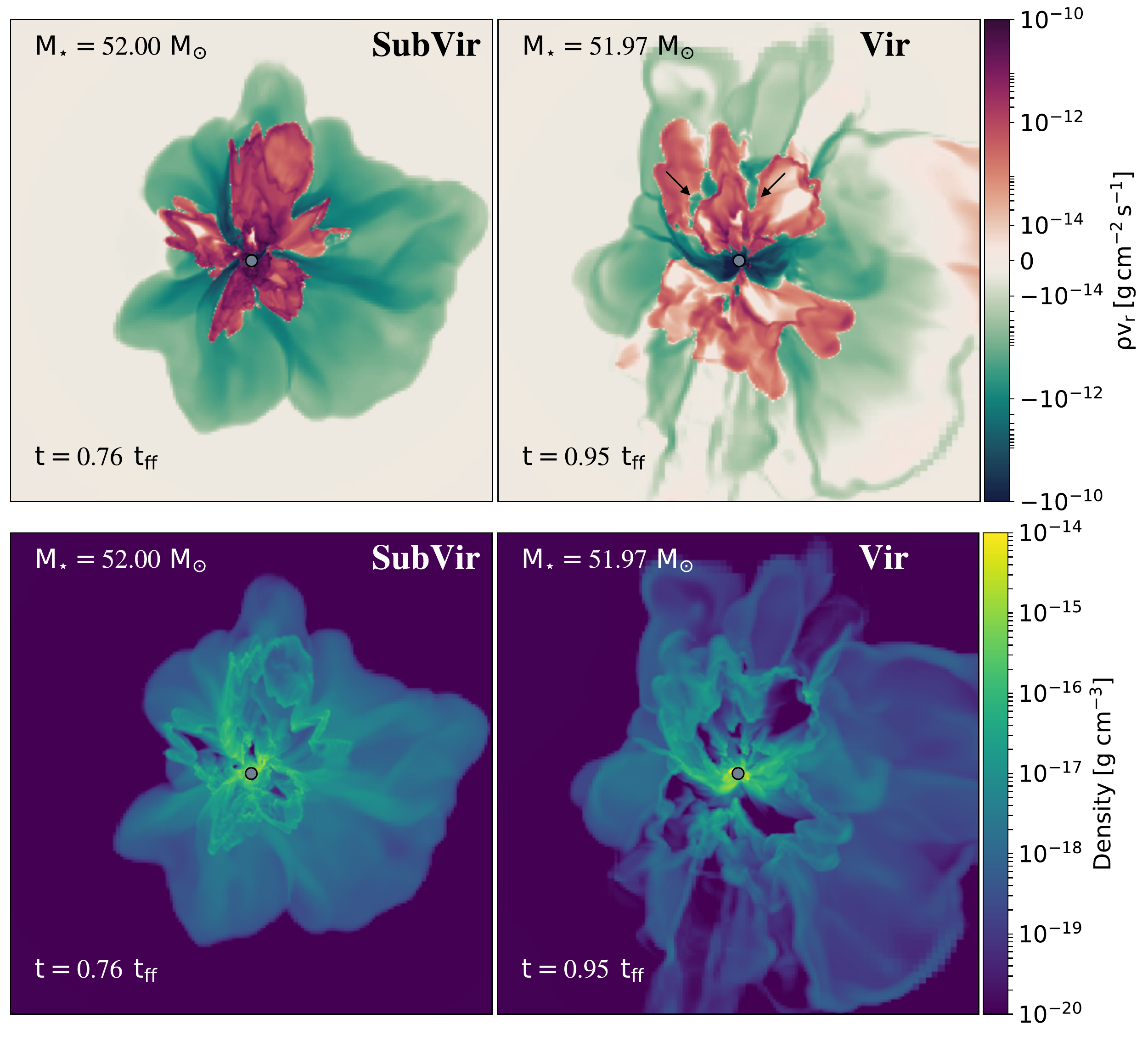}}
\caption{
\label{fig:rhovrlast}
Thin projections of the gas radial momentum, $\rho v_{\rm r}$, with respect to the massive star (top row) and gas density (bottom row) for runs \subvir\ (left panel) and \vir\ (right panel) when the most massive star has a stellar mass of $\sim52 \; M_{\rm \odot}$ corresponding to the final stellar mass in run \vir. Each panel is (50,000 AU)$^2$ in area and the projection is taken over a depth of 1000 AU with the massive star at the center. The gray \add{circle} at the center of each panel denotes the location of the massive star \addd{and the two black arrows on the top right panel denote two distinct infalling Rayleigh Taylor fingers that penetrate the expanding top bubble}.}
\end{figure}

At late times (e.g., $t=0.80 \; t_{\rm ff}$ and $t=0.87 \; t_{\rm ff}$) the average density of the radiation pressure dominated bubbles is significantly larger in run \subvir\  than the bubble density of run \vir\ in agreement with the density slice plots in Figure~\ref{fig:fedd}. This results in a larger overall radial momentum magnitude in the bubbles at late times for run \subvir\ as compared to run \vir\ as shown in Figure~\ref{fig:rhovr}. To further show this we plot the same radial momentum thin projection plots (top row) and the mass-weighted density thin projection plots (bottom row) along the $yz$-plane for both runs but at the same stellar mass at the end of run \vir, which ran to $t=0.95 \; t_{\rm ff}$ at which point the primary star has a mass of $\sim52 \; M_{\rm \odot}$ in Figure~\ref{fig:rhovrlast}. Comparing these snapshots for both runs shows that the radiation pressure dominated bubbles that are above the star are approximately the same size and that the bubbles below the primary star differ in size, with the bottom bubble in run \subvir\ being smaller than that in run \vir. This difference in bubble sizes is due to the turbulent structure and larger ram pressure of the infalling material that surrounds the bubble in run \subvir. Furthermore, the radial momentum in the radiation pressure bubbles in run \subvir\ are larger in magnitude than the bubbles in run \vir\ because the bubbles in run \subvir\ have a larger overall density as shown in the bottom row of Figure~\ref{fig:rhovrlast}. Since the overall core density in run \vir\ is smaller in magnitude than run \subvir\ as the cores evolve, radiation pressure is more efficient at launching the gas leading to lower densities in the bubbles.

\begin{figure*}
\centerline{\includegraphics[trim=0.2cm 0.2cm 0.2cm 0.2cm,clip,width=0.95\textwidth]{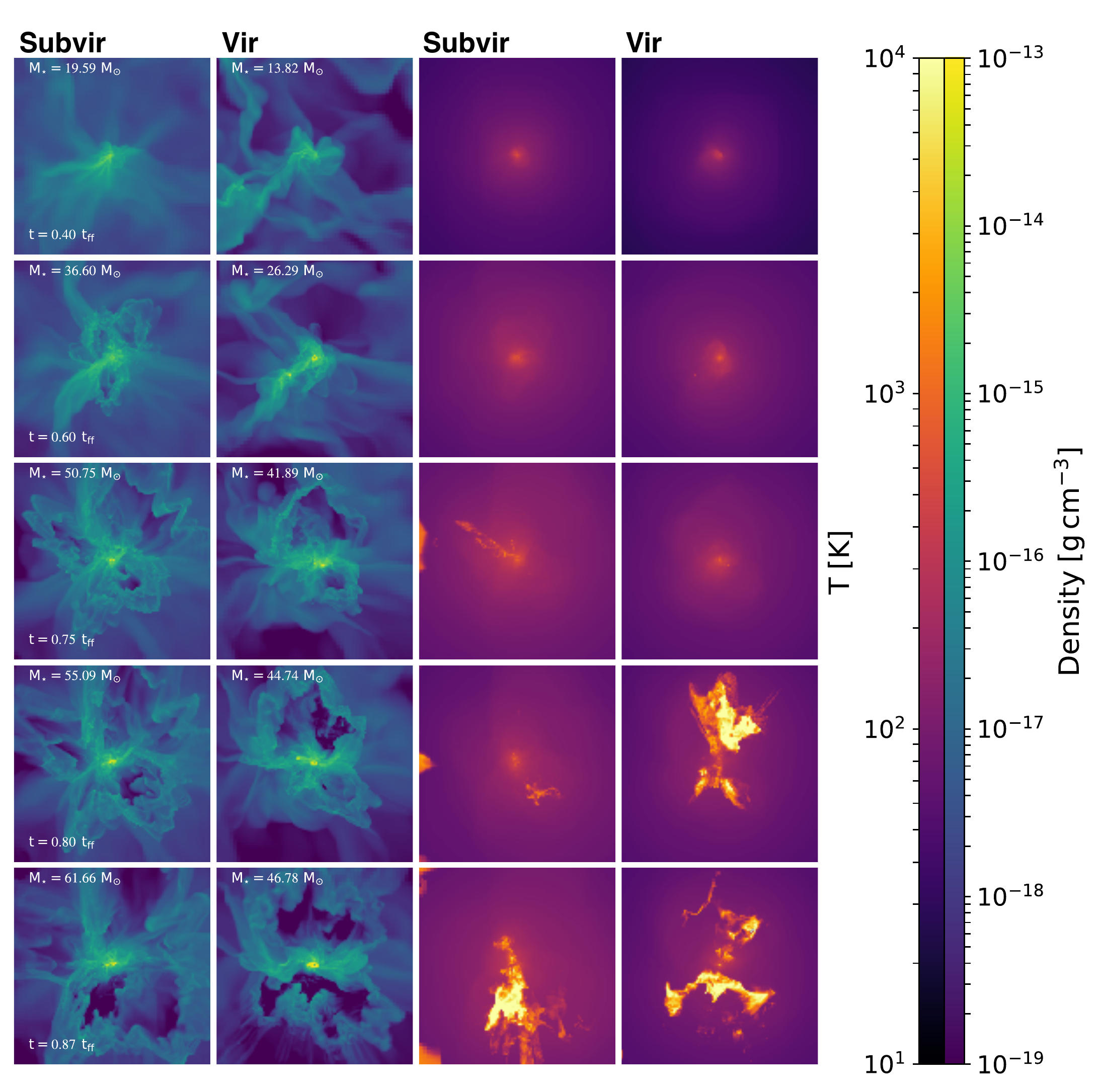}}
\caption{
\label{fig:rho_temp}
Thin density and mass-weighted temperature projections of the center of the collapsing core for runs \subvir\ (density: left most column, temperature: center right column) and \vir\ (density: middle left column, temperature: right most column). Each panel is (20,000 AU)$^2$ with the center corresponding to the location of the most massive star. The projection is taken over a depth of 1000 AU with the massive star at the center.
}
\end{figure*}

\add{We show thin projections of the mass-weighted density and temperature for runs \subvir\ (density: left hand column, temperature: middle right column) and \vir\ (density: center left column, temperature: right hand column) in Figure~\ref{fig:rho_temp} that show the growth of the radiation pressure dominated bubbles and shell structure as each simulation progresses.  Each panel is (20,000 AU)$^2$ in size and the depth of the projection is 1000 AU with the massive star at the center of each panel. The low density radiation pressure dominated bubbles eventually become transparent to the stellar radiation field since the bubbles exceed the dust sublimation temperature of 1500 K due to radiative heating from the massive star \citep{Semenov2003a}. When this occurs the majority of the stellar radiation is absorbed in the dense bubble shells and the momentum and energythereby accelerating the shells' expansion.}

\add{The second (third) panel for run \subvir\ (\vir) show that the shells of the low-density radiation pressure dominated bubbles develop small scale turbulence as they expand. This turbulence may be due to the growth of radiative Rayleigh Taylor (RT) instabilities that develop in the dense shells.} 
\addd{These instabilities occur at the interface between two fluids of different densities in which the fluid at the bottom of the interface is a lower-density radiation pressure dominated fluid (e.g., the rarefied radiation pressure dominated bubbles) that pushes on a higher density, less radiatively dominated medium \citep[e.g., the dense shells that surround these bubbles;][]{Jacquet2011a}.}
\add{In classical RT instability theory \citep[e.g.,][]{Chandrasekhar1961a}, the amplitude $\eta$ of linear perturbations that develop due to corrugations or asymmetries that form at the interface will grow with time as $\eta(t) \propto \exp{\omega t}$ where}
\begin{equation}
\label{eqn:rt}
\omega = \sqrt{g k \frac{\rho_2 - \rho_1}{\rho_2+\rho_1}} 
\end{equation}

\noindent
\add{and $\rho_1$ and $\rho_2$ are the densities at the interface (i.e., the shell and radiation dominated bubble, respectively), $g$ is the magnitude of the gravitational acceleration, k=$2\pi/\lambda$ is the wave number, and $\lambda$ is the wavelength of the perturbation \citep{Jacquet2011a}. In the limit where $\rho_2 \gg \rho_1$, such as at the interfaces between the dense shells and low-density radiatively driven bubbles presented in this work, Equation~\ref{eqn:rt} reduces to $\omega \approx \sqrt{-g k}$. Hence, the interface becomes unstable when $\rho_2 < \rho_1$.}

\add{As discussed in \pap, the initial turbulence and anisotropic absorption of the stellar radiation field seeds the development of these perturbations that can grow exponentially with time as the rarefied bubbles expand. These instabilities may grow into RT fingers, as shown in the right top panel Figure~\ref{fig:rhovrlast} that shows the radial momentum with respect to the star and shows two fingers penetrating the top bubble. Comparison with the last bottom panel in Figure~\ref{fig:rhovr} show that these fingers likely originate from the two corrugations present in the top bubble at this time. Eventually, these fingers may be able to fall to the star-disk system and deliver material to the star potentially aiding accretion} \addd{unless radiation pressure from the star overturns their infall. However, given the computational expense we are unable to follow the further evolution of the growth of these RT fingers to determine if they will be incorporated into the star-disk system.}

\subsection{Accretion History and Accretion Disk Evolution}
\label{sec:acc}
\begin{figure}
\centerline{\includegraphics[trim=0.0cm 0.2cm 0.2cm 0.2cm,clip,width=0.95\columnwidth]{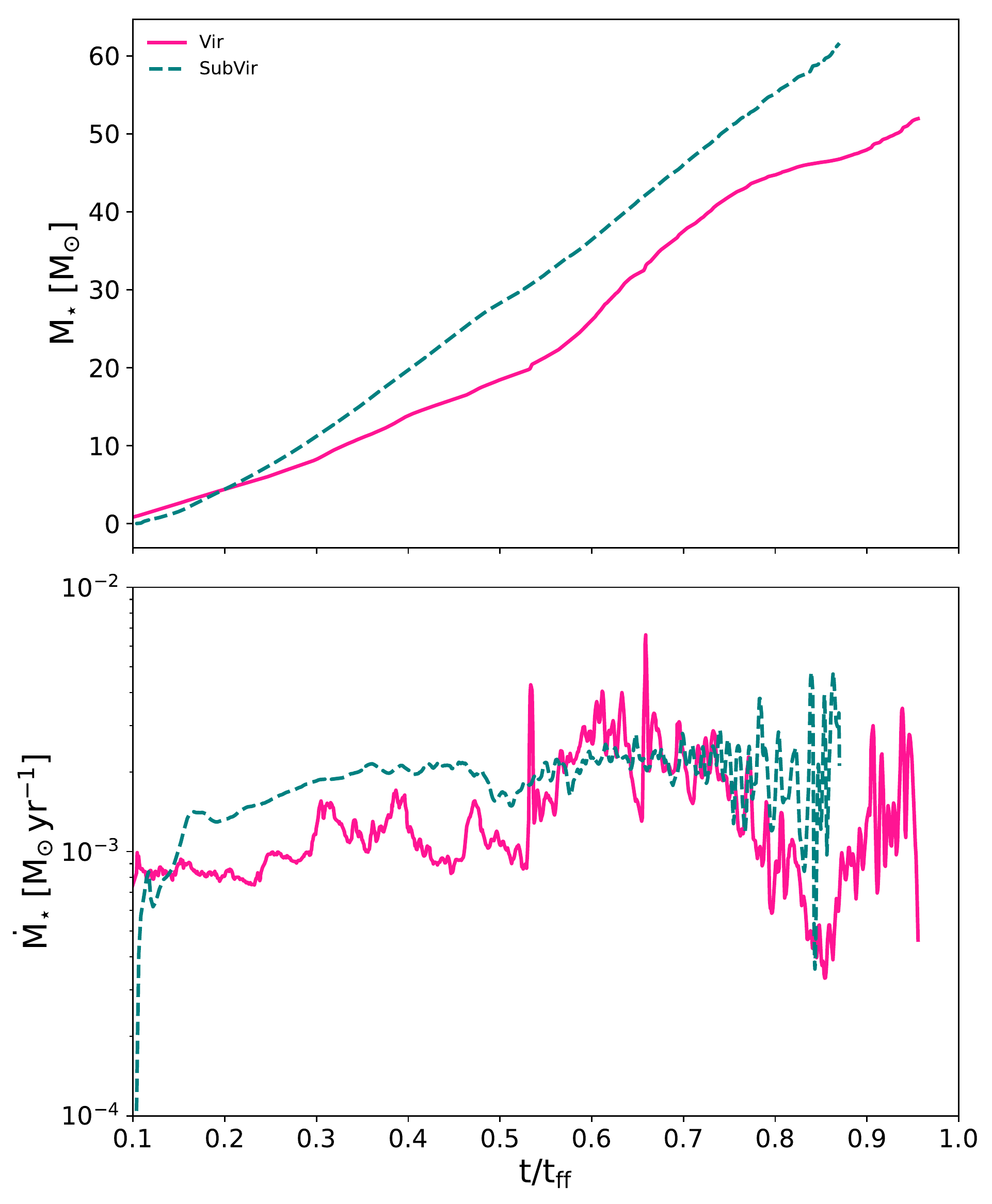}}
\caption{
\label{fig:mdott}
Primary stellar mass (top panel) and accretion rate (bottom panel) for the most massive star formed in runs \subvir\ (pink solid lines) and run \vir\ (teal dashed lines) as a function of simulation time. 
}
\end{figure}

Figure~\ref{fig:mdott} shows the stellar mass (top panel) and accretion rate (bottom panel) as a function of time for runs \subvir\ and \vir. The accretion rate for the massive star in run \subvir\ is larger by a factor of $\sim2$ as compared to the accretion rate onto the massive star in run \vir\ up to $t \approx 0.55 \; \rm{t_{\rm ff}}$, resulting in a faster stellar growth. The higher accretion rate achieved in run \subvir\ is a result of the core's global collapse as compared to run \vir, which undergoes a slower collapse due to the additional turbulent support. At $t \approx 0.55 \; t_{\rm ff}$ the accretion rate onto the massive star in run \vir\ increases and becomes comparable to the accretion rate attained in run \subvir\ and they remain comparable until $t \approx 0.75 \; t_{\rm ff}$.

Collapsing optically thick filaments inherent to the core's turbulent structure can lead to variable accretion onto the star and also overcome the radiation pressure barrier, delivering material to the star that would otherwise be blown away. In this scenario, the radiative flux escapes through low density channels while the denser filaments can accrete onto the star \citep{Rosen2016a, Goddi2018a}. As discussed in the previous Section, the filaments in run \subvir\ are wider than those that form in run \vir\ and begin to collapse instantly. This results in an accretion rate onto the primary star that grows in time and is relatively smooth up to $t \approx 0.55 \; \rm{t_{\rm ff}}$. For run \vir, the turbulence has more time to develop thin, dense filaments before they collapse onto the star as compared to run \subvir\ and the accretion of these filaments causes the variability of the accretion rate onto the star at early times (i.e., between $t=0.3 - 0.55 \; t_{\rm ff}$). This variability can increase the accretion rate by up to a factor $\sim$2 for $t \lesssim 0.55 \; t_{\rm ff}$. 

\begin{figure*}
\centerline{\includegraphics[trim=0.0cm 0.0cm 0.0cm 0.0cm,clip,width=0.9\textwidth]{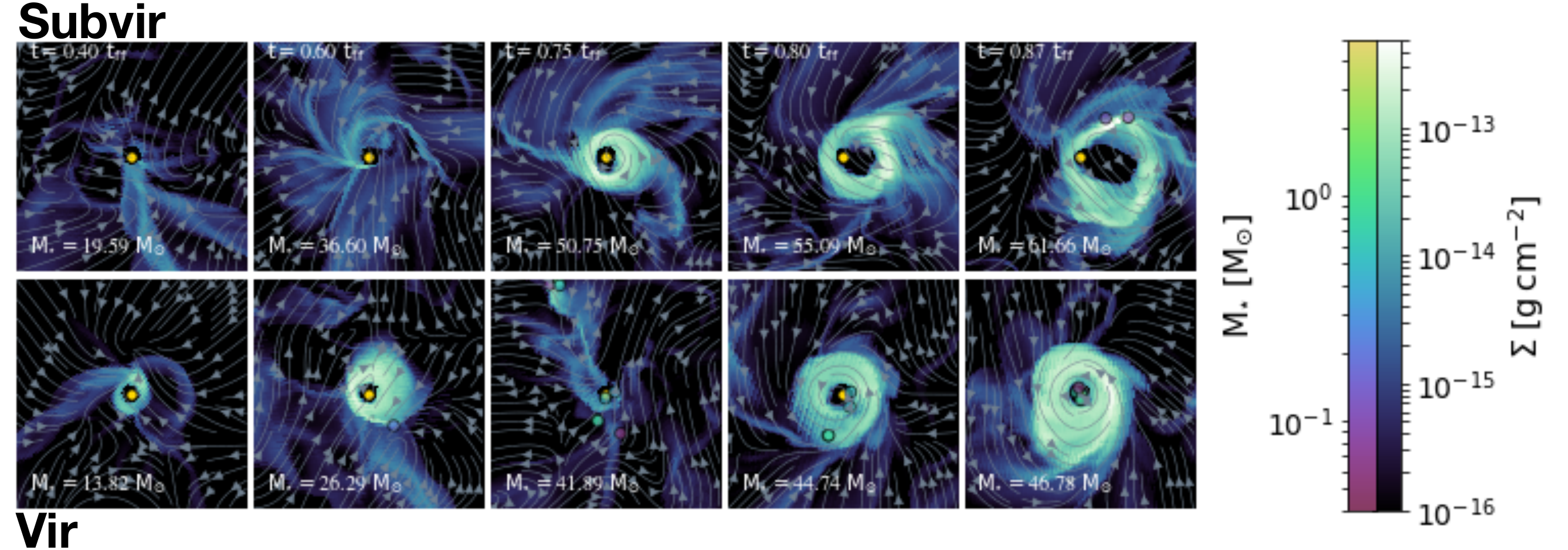}}
\caption{
\label{fig:disk}
\add{Mid-place density slices} of the accretion disk in run  \subvir\  (top row) and \vir\ (bottom row) showing the disk's time evolution. Each panel represents a projection of the accretion disk, with the most massive star at the center of the panel (yellow circle), that is (\blue{2}000 AU)$^2$ in size. \add{Velocity streamlines and }
	\add{c}ompanion stars with masses greater than 0.04 $M_{\rm \odot}$ are over-plotted on all panels.
}
\end{figure*}

At later times the accretion rate becomes much more highly variable and can quickly increase by up to an order of magnitude before dropping to lower values for both runs. This increase in accretion variability is due to accretion of infalling filaments and/or the inflow of mass from gravitational torques induced in the accretion disk. In order to determine which process dominates the accretion variability we examine the evolution and structure of the accretion disk for runs \subvir\ and \vir. Figure~\ref{fig:disk} shows a series of \add{density slices} of the accretion disk that \add{forms around} the massive star for run \subvir\ (top row) and run \vir\ (bottom row) \add{with the velocity streamlines over-plotted showing that this over-dense structure eventually undergoes circular motion around the central star}. The massive star is at the center of each panel, denoted by the \add{yellow} circle and companion stars that are color-coded by mass are over-plotted. These stars are formed via turbulent fragmentation or via disk fragmentation, which we will describe in more detail in Section~\ref{sec:frag}. As these panels show a noticeable \add{high density} accretion disk (i.e., an accretion disk with a radius larger than the 80 AU accretion zone radius of the sink particle \add{that has a circular structure and undergoes circular motion around the massive star}) begins to form around the massive star in run \vir\  at $t \approx 0.4 \; t_{\rm ff}$ whereas in run \subvir\ a noticeable accretion disk does not begin to form until $t \approx 0.65 \; t_{\rm ff}$ when the star has a mass of $41.1 \; M_{\rm \odot}$. The accretion disk grows in size as both simulations progress due to conservation of angular momentum: as the core collapses, material that is farther away from the massive star has a larger net angular momentum and therefore will be circularized at a distance farther from the star. The disk can be disrupted due to the passage of companion stars as seen in the third snapshot of the accretion disk projection for run \vir\ in Figure~\ref{fig:disk}. However, after this occurs and the core continues to collapse, infalling material is circularized rebuilding the disk.

As the accretion disk grows in size and mass, it become\add{s} gravitationally unstable \add{and develops spiral structure.} The gravitational torques induced deliver\add{s} material to the star \citep{Kratter2008a}. This, in turn, leads to a highly variable accretion rate as shown in the bottom panel of Figure~\ref{fig:mdott}. Indeed, we see that the accretion rate, while relatively smooth throughout most of the simulation becomes highly variable for run \subvir\ for $t \gtrsim 0.7 t_{\rm ff}$ and the magnitude of the variability increases with time as the both the accretion disk and massive star increase in mass. \add{This variability in run \subvir\ occurs because} the star and disk are orbiting around the star-disk system's center of mass and this further drives gravitational torques in the disk that eventually lead to disk fragmentation into companion stars. Likewise, we see the same behavior for run \vir\ but this behavior begins at earlier times, $t \sim 0.5 t_{\rm ff}$, since the accretion disk forms earlier than the disk in run \subvir. Hence, we find that at late times for both runs most of the accretion variability is due to how mass is driven from the disk to the star via gravitational torques. 


\begin{figure}
\centerline{\includegraphics[trim=0.0cm 0.2cm 0.2cm 0.2cm,clip,width=1\columnwidth]{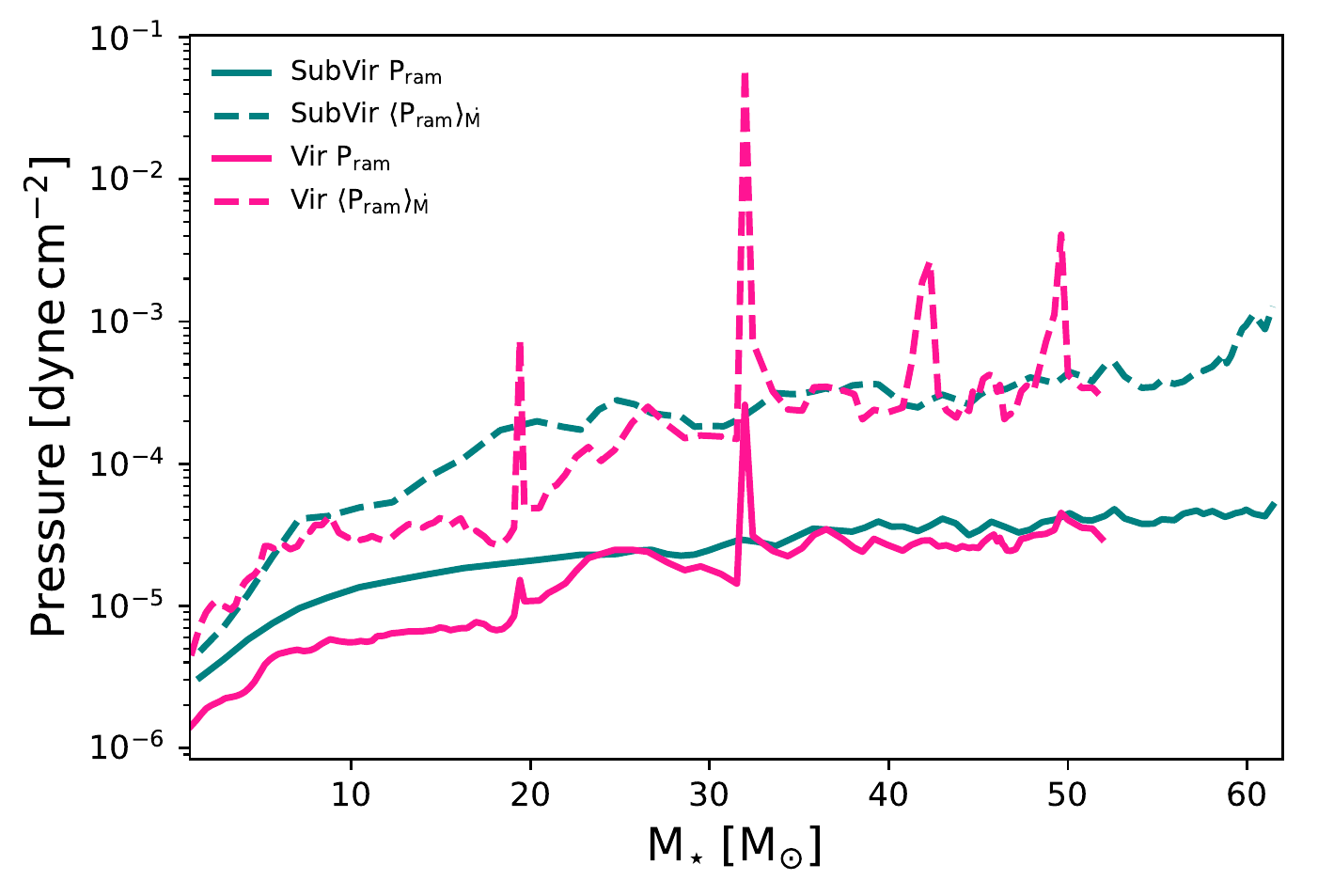}}
\caption{
\label{fig:pram}
Comparison of the area-weighted (solid lines) and mass-weighted ram pressure (dashed lines) for run \subvir\ (teal lines)  and run \vir\ (pink lines) from inflowing material taken over a 1000 AU sphere surrounding the accreting primary star as a function of stellar mass. See main text for full details on how these averages are defined.
}
\end{figure}

While most of the accretion variability is due to gravitational torques in the accretion disk driving material to the central massive star at late times, accretion variability is also a result of infalling filaments and RT instabilities that accrete onto the star-disk system. To determine this and differentiate it from accretion variability attributed to the accretion disk we plot the ram pressure of the inflowing core material in Figure~\ref{fig:pram} for material that is 1000 AU away from the star as a function of primary stellar mass since the accretion disk radius is $<$ 1000 AU throughout both simulations. To compute the inflow ram pressure we take a sphere 1000 AU in radius centered on the most massive star for each run. We compute the area-weighted mean ram pressure (solid lines) and the mass-flux-weighted mean ram pressure (dashed lines) for inflowing material for runs \subvir\ (teal lines) and \vir\ (pink lines). These quantities are defined by 
\begin{equation}
\langle P_{\rm ram}\rangle_w = \frac{\int \rho v_{\rm r}^2 w\rm{d}A}{\int w \rm{d}A}
\end{equation}
where $v_{\rm r}$ is the radial velocity, and the weighting function $w$ is unity for the area-weighted average($\langle P_{\rm ram} \rangle$), and we take $w=\rho v_{\rm r}$ for the mass-flux-weighted average ($\langle P_{\rm ram} \rangle_{\dot{\rm M}}$) where we only include contributions from gas that is inflowing to the star (i.e., $v_{\rm r}>0$). We include the mass-flux-weighted mean ram pressure because it is a better representation of the ram pressure of the material that can be accreted onto the star. 

Throughout the simulation, the ram pressure of the inflowing material is larger for run \subvir\ except at points when the accretion rate onto the star jumps significantly for run \vir. These jumps in ram pressure are much more pronounced in the mass-flux-weighted mean ram pressure and can cause it to jump by up to $\sim 3$ orders of magnitude. By comparing these ram pressure spikes in run \vir\ with the accretion rate shown in the bottom panel of Figure~\ref{fig:mdott}, we find that these spikes correspond to large jumps in the accretion rate. Hence, we find that the turbulent structure of the core can lead to highly variable accretion onto the massive star for run \vir\ but is negligible for run \subvir. This is likely because the core in run \vir\ develops a more filamentary structure than the core in run \subvir\ because the core does not undergo instantaneous collapse and therefore has more time for supersonic shocks to collide to form more thin, dense filaments.

\subsection{Core and Disk Fragmentation into Companion Stars}
\label{sec:frag} 
\begin{figure}
\centerline{\includegraphics[trim=0.0cm 0.2cm 0.2cm 0.2cm,clip,width=1.0\columnwidth]{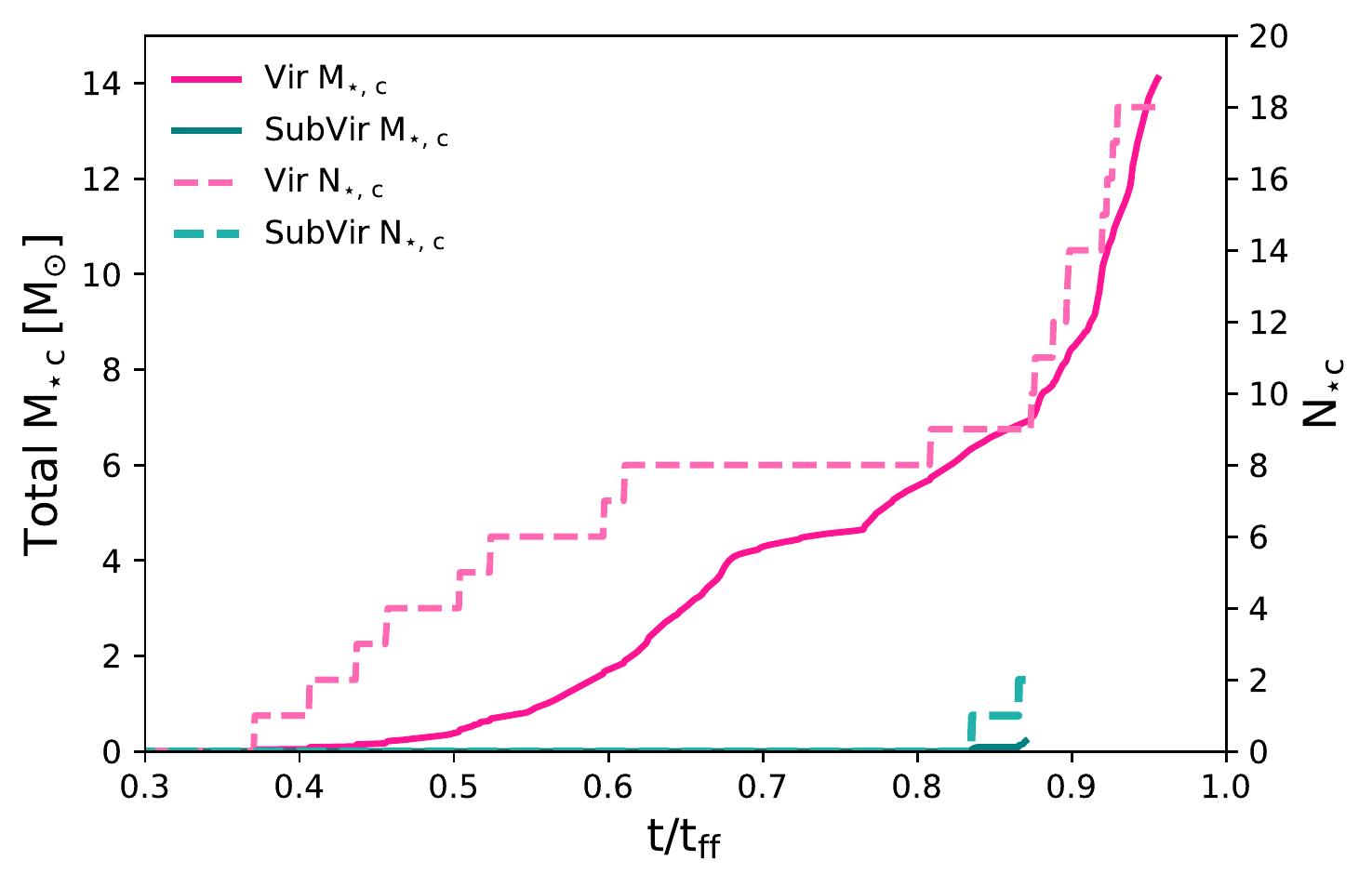}}
\caption{
\label{fig:mcnctot}
Total mass and number of companion stars as a function of time for runs \subvir\ and \vir. The solid lines (dashed lines) show the total companion stellar mass (number of companion stars) for runs \subvir\ (teal lines) and \vir\ (pink lines).
}
\end{figure}

Companion stars can form via turbulent fragmentation or disk fragmentation. Turbulent fragmentation occurs in molecular clouds when supersonic shocks collide and form high density thin sheets and elongated structures surrounded by low-density voids (i.e., filaments). Dense parts of these structures fragment into dense clumps and cores when they become Jeans' unstable \citep{Padoan2002a}. The hierarchical fragmentation continues until stars form inside the dense cores. Since the filaments are due to the collision of supersonic shocks in the ISM, the degree of fragmentation depends on the velocity dispersion and therefore the virial state of the core. We find that the core in run \vir\ undergoes significant turbulent fragmentation but the core in run \subvir\ does not undergo any turbulent fragmentation as shown in Figure~\ref{fig:mcnctot} that shows the total mass and number of companion stars as a function of time for runs \subvir\ (teal lines) and \vir\ (pink lines). The pre-stellar core in run \vir\ has a larger initial velocity dispersion than the core in run \subvir\ and therefore provides support against collapse allowing time for the filaments to form and then undergo longitudinal collapse (i.e., collapse modes that are parallel to the symmetry axis of the filament) due to their self-gravity \citep{Bastian1983a}.  In contrast, the core in run \subvir\ has less supersonic shocks due to its weaker turbulence and lower $\mathcal{M}$.  Because of these effects and aided by the fact that the core in run \subvir\ instantly collapses,  high-density filaments that form do not have enough time to develop thin, elongated structures that can then collapse into stars since their motions are dominated by infall towards the primary star. Therefore, they do not become Jeans unstable and fragment into low-mass companion stars. 

As Figure~\ref{fig:mcnctot} shows, the core in run \vir\ starts to fragment into companion stars via turbulent fragmentation at $t \approx 0.35 \; t_{\rm ff}$, forming eight companion stars by $t \approx 0.60 \; t_{\rm ff}$. After that, turbulent fragmentation into companion stars for run \vir\ is suppressed. This is likely due to the enhanced radiative heating of the core material by the massive star when the star reaches $\sim 25 M_{\rm \odot}$, which we describe in more detail in the next subsection. At $t \approx 0.87 \; t_{\rm ff}$ the formation rate of companion stars is greatly enhanced due to disk fragmentation, nearly doubling the number of companion stars by a factor of two by $t \approx 0.9 \; t_{\rm ff}$. This occurs because the disk becomes gravitationally unstable because the massive star and disk are orbiting their common center of mass. The gravitational torques induced in the disk drive high density spiral waves that can become Jeans' unstable and collapse to form disk borne stars \citep{Kratter2006a}. Referring back to Figure~\ref{fig:disk} and Figure~\ref{fig:mcnctot}, we see that the accretion disk in run \subvir\ begins to fragment at the end of the simulation, forming two companion stars with $M_{\rm \star} > 0.04 \; M_{\rm \odot}$ (i.e., the mass that we initialize our protostellar model). 

\begin{figure}
\centerline{\includegraphics[trim=0.2cm 0.2cm 0.2cm 0.2cm,clip,width=0.9\columnwidth]{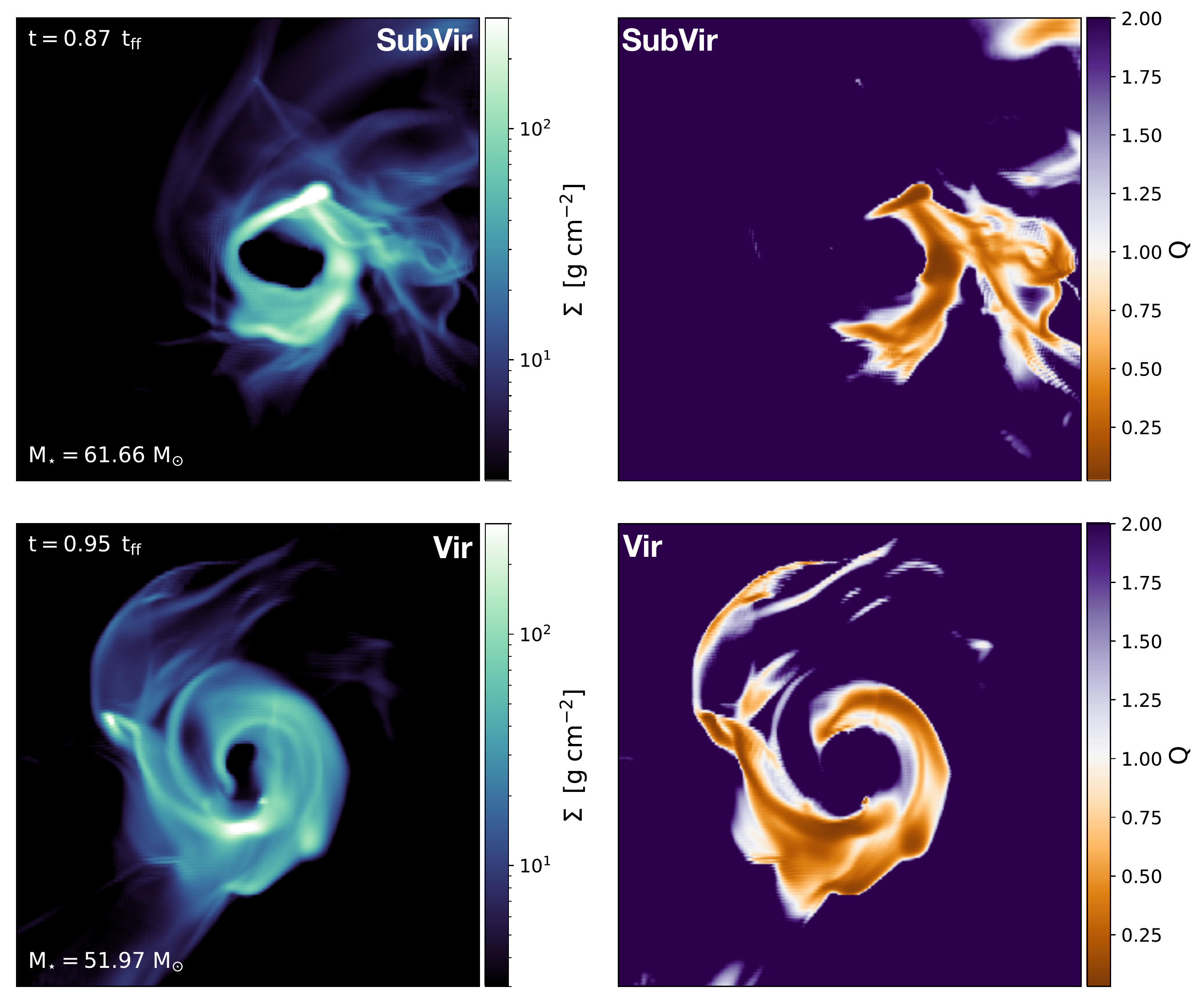}}
\caption{
\label{fig:disklast}
\add{Final snapshot of the surface density (left panels) and Toomre $Q$ parameter (right panels) of the accretion disk in runs \subvir\ (top row) and \vir\ (bottom row). Each panel is (3000 AU)$^2$ in size. The projection is taken over a height of 1000 AU above and below the massive star and the massive star is located at the center of each panel. Regions of the disk with $Q < 1$ denote locations that are unstable to gravitational collapse.
}}
\end{figure}

\add{To demonstrate how the accretion disk becomes unstable and fragments to form close-in companion stars we show the disk surface density (left panels) and disk Toomre Q parameter (right panels) for the final snapshots of runs \subvir\ (top row) and \vir\ (bottom row) in Figure~\ref{fig:disklast}. The Toomre Q parameter is given by
\begin{equation}
Q = \frac{c_s \kappa}{\pi G \Sigma}
\end{equation}
where $c_{\rm s}$ is the sound speed, $\kappa$ is the epicyclic frequency which is equal to the angular velocity $\Omega$ for a Keplerian disk, and $\Sigma$ is the disk surface density \citep{Toomre1964a}. Values of $Q<1$ denote locations where the disk is unstable to gravitational collapse whereas regions with $Q>1$ are stable and not prone to collapse.} For run \vir, \add{the majority of the disk has $Q<1$} and the disk \add{has} fragment\add{ed} into several low-mass companion stars \add{by this time. We also note that the final disks in both simulations are relatively massive with masses of $\sim 10.5 \; M_{\rm \odot}$ for run \subvir\  and $\sim9.4 \; M_{\rm \odot}$ for run \vir\ and they are still being fed by the collapsing core at this time.}

\begin{figure}
\centerline{\includegraphics[trim=0.0cm 0.2cm 0.2cm 0.2cm,clip,width=0.9\columnwidth]{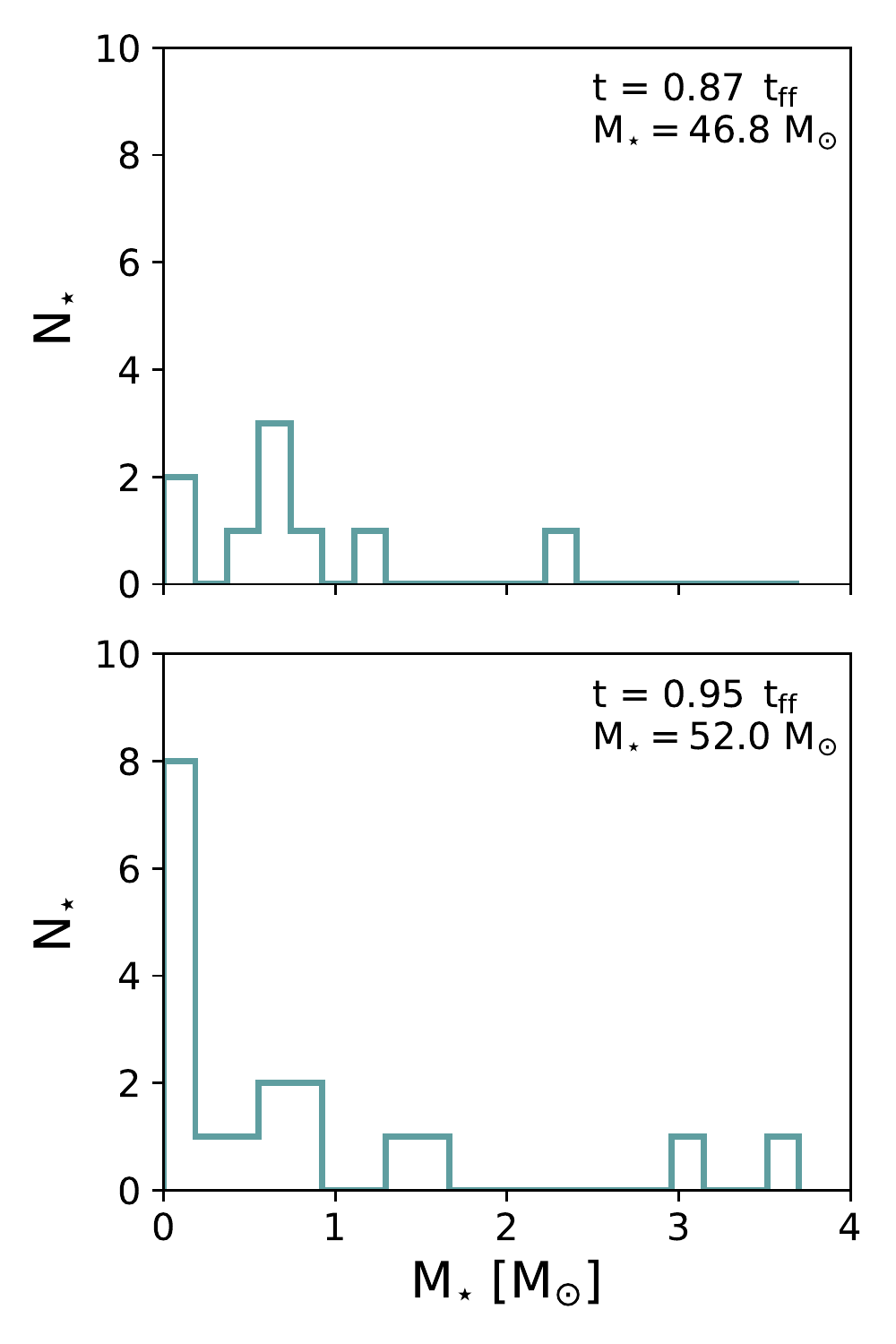}}
\caption{
\label{fig:nchist}
Histogram of the stellar masses of the companion stars that form in run \vir\ right before the disk undergoes significant fragmentation  (top panel) and at the end of the simulation (bottom panel). The top right corner of each panel lists the time and primary stellar mass.
}
\end{figure}

Most of the disk-borne stars \add{in run \vir\ }are very low-mass, as shown in the histogram in Figure~\ref{fig:nchist} that shows the distribution of stellar masses for the companion stars before disk fragmentation (top panel) compared to the final distribution of companion \add{star} masses at the end of the simulation (bottom panel). The bottom panel shows that the newly formed disk borne stars have masses $\lesssim 0.2 \; M_{\rm \odot}$ whereas several of the stars that formed via turbulent fragmentation have attained masses $\gtrsim 1\;  M_{\rm \odot}$ by the end of the simulation. At the end of both simulations we find that they consist of a hierarchical system that contains a massive primary and a series of low-mass companions instead of a massive binary system. This may also be attributed to our treatment of merging sink particles because we do not merge sink particles once their mass exceeds $0.04 \; M_{\rm \odot}$ and therefore our strict merging criteria may lead to an excess of low-mass companions, which we discuss further in Section~\ref{sec:disc}.

\subsection{Evolution of the Core Properties}
\label{sec:coreprop}
\begin{figure*}
\centerline{\includegraphics[trim=0.2cm 0.0cm 0.2cm 0.2cm,clip,width=0.85\textwidth]{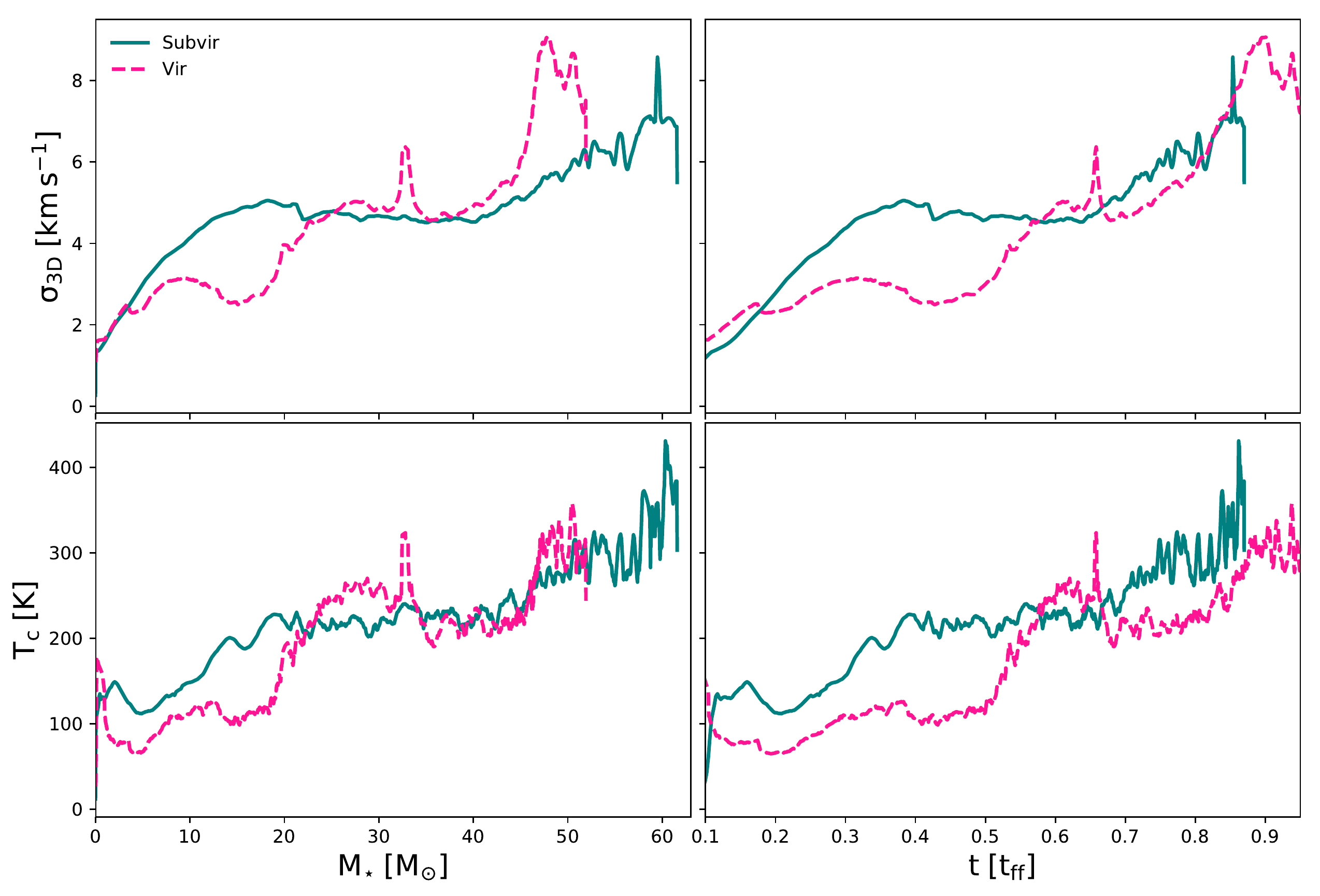}}
\caption{
\label{fig:cloud_props}
Mass-weighted 3D velocity dispersion ($\sigma_{\rm 3D}$; top row) and temperature ($T_{\rm c}$; bottom row) of the core in runs \subvir\ (teal solid lines) and \vir\ (pink dashed lines) as a function of primary stellar mass (left column) and simulation time (right column).
}
\end{figure*}

We show the average mass-weighted core velocity dispersion (top row) and temperature (bottom row) as a function of primary stellar mass (left column) and time (right column) for runs \subvir\ (teal solid lines) and \vir\ (pink dashed lines) in Figure~\ref{fig:cloud_props}, respectively. We plot these values as a function of both simulation time and primary stellar mass because time is a good proxy to follow these properties as the core collapses whereas primary stellar mass traces the impact of stellar feedback. As \citet{Traficante2018a} noted infall motions leads to an increase in the velocity dispersion in collapsing clouds. In agreement, we see that the rapid collapse of the core in run \subvir\ causes the core's \add{mass-weighted} velocity dispersion to be\add{come} larger than the \add{mass-weighted} velocity dispersion in run \vir\ \add{beginning} at $t \approx 0.2 \; t_{\rm ff}$ when the primary star is $\sim$few $M_{\rm \odot}$ such that radiative feedback is unimportant at this time. The velocity dispersion for run \subvir\ continues to increase and remains larger than the velocity dispersion in run \vir\ until the star has a reached a mass of $\sim25 \; M_{\rm \odot}$. \add{At this stellar mass for both runs radiative feedback from the massive star becomes super-Eddington and launches radiation pressure dominated bubbles along the polar directions of the star. As these bubbles expand they become RT unstable and drive turbulence in the shells as shown in the thin mass-weighted density projections for runs \subvir\ (left most column) and \vir\ (center left column) in Figure~\ref{fig:rho_temp} \citep[e.g.,][]{Krumholz2012b}, thereby increasing the overall core's velocity dispersion in both runs.} Hence, we find that both gravitational infall and radiative feedback can lead to an increase in the overall velocity dispersion of the cores.

Similarly, gravitational collapse and radiative feedback also increases the global core temperature as shown in the \add{bottom panels} of Figure~\ref{fig:cloud_props}. As the cores in runs \subvir\ and \vir\ collapse and the primary stars grow in mass their overall mass-weighted temperature increases with time, however we see that this effect is much more pronounced for run \subvir\ up to $t \approx 0.6 \; t_{\rm ff}$ (corresponding to $M_{\rm \star} \sim 25 \; M_{\rm \odot}$ for the primary star in run \subvir) in agreement with the behavior seen in the core's velocity dispersion. As described in Section~\ref{sec:frag}, turbulent fragmentation for run \vir\ occurs during $t \approx 0.35-0.6 \; t_{\rm ff}$ but no turbulent fragmentation occurs for run \subvir\ during this time. Radiative heating reduces fragmentation and therefore we find that both radiative heating from the more massive star in run \subvir\ and enhanced heating due to collapse also suppresses fragmentation of the core in run \subvir\ as compared to run \vir\ \citep{Offner2009a}. As the primary star becomes more massive, making radiative heating more important, the core temperature continues to increase for both runs. However, radiative feedback is more efficient at heating the core for run \vir\ when the star has a mass of $\sim$25 to $\sim$37 $M_{\rm \odot}$. At higher stellar masses the temperature is comparable for both simulations. Hence, we find that at early times gravitational collapse can lead to higher core temperatures whereas at late times and high stellar masses the overall \add{increase in the} core temperature is \add{due to} stellar feedback.

\section{Discussion}
\label{sec:disc}
The purpose of this work is to understand how the initial dynamic state of the core affects the accretion history of massive stars, core fragmentation, and formation of companion stars. 
Most notably we find that highly subvirial pre-stellar cores undergo rapid, monolithic collapse and no turbulent fragmentation whereas virialized cores undergo a slower, more gradual collapse and significant fragmentation thus forming a small cluster of stars. In what follows, we discuss how the pre-stellar core's virial state affects the growth rate of massive stars and core fragmentation in Section~\ref{sec:form} and how disk fragmentation can produce close-in companion stars in Section~\ref{sec:diskfrag}. The simulations presented here do not include magnetic fields which are ubiquitous  in massive star forming regions and are therefore an important ingredient in the star formation process. In light of this, we discuss our motivation for neglecting them, what effect we expect magnetic fields to play in the collapse of massive cores into massive stellar systems, and our plans to include them and other forms of stellar feedback such as collimated outflows in future work in Section~\ref{sec:caveats}.

\subsection{Fragmentation and Multiplicity}
\subsubsection{Formation of Isolated Massive Stars versus Protoclusters via Turbulent Fragmentation}
\label{sec:form}
In this work we showed that the dynamic state of massive pre-stellar cores, as described by their virial parameter that compares the core's kinetic energy to its gravitational potential energy, affects the accretion history of massive stars and core fragmentation into companion stars. Specifically, we find that virialized cores for the initial conditions chosen in this work undergo hierarchical turbulent fragmentation up until $t\sim 0.6 \; t_{\rm ff}$, when radiative heating from the massive star becomes significant. This results in a small cluster that contains one massive star with several low-mass companions. In contrast, we find that highly subvirial cores undergo rapid, monolithic collapse that yields higher accretion rates onto massive stars and no turbulent core fragmentation into companion stars. \add{When the star is sufficiently massive} the enhanced radiative heating, in comparison to run \vir, heats up the core material and therefore suppresses fragmentation. Therefore, we conclude that subvirial cores are more likely to form massive stars \add{with higher accretion rates at the onset of core collapse} than virialized cores that have \add{similar} physical properties (e.g., surface density, mass, and density profile) \add{and that the lower degree of turbulence in subvirial cores likely result in less turbulent fragmentation for subvirial cores}. Our results suggest that star clusters that host at least one massive star \add{may have} originated from cores that have a greater degree of turbulence at the onset of collapse and that \addd{wide} companion stars form either coevally or at a later stage after the birth of massive protostars in agreement with observations \citep{Zhang2015a, Pillai2019a}.

Similarly, \citet{Fontani2018a} studied the early collapse of massive supercritical magnetized cores with different initial virial parameters and magnetic field strengths by performing a suite of radiation-magnetohydrodynamic simulations and found that fragmentation is inhibited for clumps with low turbulence ($\mathcal{M} \lesssim 3$) regardless of the magnetic field strength in agreement with run \subvir\ presented in this work ($\mathcal{M} = 1.7$). They conclude that magnetized cores with a high initial turbulence ($\mathcal{M} \sim 6$) undergo fragmentation but that the number of fragments that form is larger for weakly magnetized cores. They conclude that the turbulent state of the core is more important than magnetic fields in determining whether the core will fragment into a small cluster or not but that the degree of fragmentation and distribution of fragments does depend on the magnetic field strength. Similarly, in the absence of magnetic fields we find that significant fragmentation also occurs in massive cores with weaker initial turbulence ($\mathcal{M} \sim 4.5$, run \vir\ presented in this work) leading to the formation of a small cluster in which $\sim 9$ low-mass companion stars are formed via turbulent fragmentation. 

The simulations presented here and in \citet{Fontani2018a} only explore a small parameter space of massive pre-stellar cores, however observed massive dense cores span a wide range of physical properties. Based on our results that the initial turbulence of these cores affect fragmentation we expect that weakly subvirial cores with $\alpha_{\rm} > 0.14$, the initial value used for run \subvir, should undergo fragmentation but to a lesser degree than virialized cores. In agreement, observations of massive star-forming cores show a diverse population of fragments. \citet{Beuther2018b} studied 20 evolved high-mass star-forming regions that host at least one young massive stellar object (MYSO; i.e., $M_{\rm \star} \gtrsim 8 \; M_{\rm \odot}$) and found that these clumps have diverse fragmentation morphologies ranging from regions that are dominated by single evolved high-mass cores to those that fragment up to 20 cores. They conclude that the diversity in fragmentation of evolved massive-star forming clumps depends on the clump's dynamic state, which depends on the interplay of self-gravity, turbulence, magnetic fields, and heating from stellar feedback that can suppress fragmentation. Hence, one way to possibly identify massive stars that form from highly subvirial cores would be to find young, embedded massive stars that have no long distance companions but \add{close-in} companions due to disk fragmentation (\citet{Kratter2006a}, see Section~\ref{sec:diskfrag}). 
 
\subsubsection{Formation of Close-In Companion Stars via Disk Fragmentation}
\label{sec:diskfrag}

Turbulent fragmentation is likely responsible for wide binaries but $\sim80 \%$ of O-stars ($M_{\rm \star} \gtrsim 16 \; M_{\rm \odot}$) are found in close-in multiple binaries with separations $\lesssim 700$ AU  \add{and a flat mass-ratio, $q=M_{\rm 2}/M_{\rm 1},$ distribution} where $M_{\rm 1}$ ($M_{\rm 2}$) is the primary (secondary) stellar mass \citep{Sana2012a, Sana2014a}. This large binary fraction may originate from the formation process via disk fragmentation rather than by direct capture \citep{Kratter2006a}. In agreement with this scenario, we find close-in companion stars form via disk fragmentation at late times regardless of the initial virial state of the core.  We find that the companion stars formed via disk fragmentation at the end of the simulations presented in this work have $M_{\rm 2} \ll M_{\rm 1}$ resulting in extreme mass ratios. 

In agreement with our results, recent observational studies have now found asymmetric, fragmented Keplerian accretion disks around massive proto-O stars. \citet{Zapata2019a} found a binary system of compact dusty objects that are separated by $\approx$300 AU in projection at the center of an asymmetric accretion disk of size $\approx 1200$ AU. They conclude that the primary massive proto-O star has a stellar mass of $\sim 20 \; M_{\rm \odot}$ and its low-mass companion likely has a mass within $0.2-2 \; M_{\rm \odot}$ yielding a mass ratio of $q \approx 0.01-0.1$. Similarly, \citet{Ilee2018b} observed the G11.92-0.61 system with ALMA  and found a fragmented disk around a $\sim34 \pm 5 \; M_{\rm \odot}$ proto-O star with a low-mass companion star with mass $\lesssim 0.6 \; M_{\rm \odot}$ yielding a mass ratio of $q \sim 0.015$. These studies are the first to observe the formation of a binary star via disk fragmentation around a young proto O-star. 

The extreme mass ratio observed in these objects suggest that disk fragmentation into low-mass companions \add{can} occur when the primary star is substantially massive in agreement with the simulations presented in this work and numerical work by \citet{Meyer2018a}. As these systems evolve, the secondary can become more massive by accreting disk material or merge with other fragments so that it becomes massive enough to resist being dragged inward by the primary star. Further accretion onto the secondary, if it begins to grow faster than the primary, can then push the mass-ratio from an extremely low-value at the onset of disk fragmentation to $q \approx 1$ at late times after the core has been exhausted \citep{Krumholz2009a} consistent with observations \citep[\add{e.g.,}][]{Chini2012a, Sana2014a, Pomohaci2019a}. \add{For example, \citet{Duffell2019a} find that binary systems with initial extreme mass ratios, $q \ll 1$ can be pushed to higher values of $q$ as the binary system accretes material from the disk because the accretion onto the secondary can be as much as a factor of $10$ larger than the primary star's accretion rate pushing the system to a higher mass ratio.} However, we are unable to follow the long-term growth of the low-mass companion stars \add{or further disk fragmentation into additional companion stars} due to the high computational cost\add{\footnote{\add{The simulations presented in this work each took $\approx$ few $ \times10^5$ CPU hours to run, with a significant fraction of the time spent on the final few kyr. This is because at late times the intense luminosity from the primary star yields such strong radiative accelerations such that the timestep, set by the Courant condition, decreases drastically thereby making advancing the simulation prohibitively expensive.}}} of the simulations presented in this work. 

Additionally, the growth of disk-borne stars is also sensitive to our merging criteria for star particles. In this work we only allow two sink particles to merge when the lower mass particle has a mass less than 0.04 $M_{\rm \odot}$. Hence, this strict criterion for particle merging may lead to an over abundance of very low-mass companion stars as discussed in \pap. \add{This assumption in merging sink particles may lead us to predict a higher multiplicity for the massive system formed in run \vir\ as compared to \subvir\ because we see that in Figure~\ref{fig:disk}, many of the low-mass companion stars that are formed in the disk at late times would merge if our merging criteria had no mass dependence.} \addd{A more lenient prescription of our merging criteria such as allowing sinks to merge if they pass within one accretion radius of each other (80 AU), as was done in \citet{Krumholz2009a}, would lead to a lower number of higher mass companion stars that may be more consistent with observations of massive multiple stellar systems. In light of these limitations, we are unable to conclude if such massive multiple stellar systems are formed quickly via disk fragmentation and accretion or via capture due to dynamical interactions in clustered environments.} 

\addd{In the simulations presented here we treat the stars as accreting sink particles with an 80 AU accretion zone and therefore we are not able to follow if such objects are disrupted via shear motions on smaller scales in the accretion disk. For comparison, \citet{Meyer2018a} performed similar simulations of disk fragmentation around massive protostars to study disk fragmentation with a much higher spatial resolution. By using a spherically symmetric domain in which the resolution increases logarithmically from the origin they attain sub-au resolution closest to the massive star and $< 10$ AU  resolution in the $\sim 500$ AU region of their domain. They find that fragments that form via gravitational instability are not disrupted by shear motions in the accretion disk at these size scales and continue to grow via accretion like the companion stars formed in this work. Hence, we expect that the companion stars that form via disk fragmentation will likely survive the shear forces in the accretion disk unless the fragments, before sink creation, is significantly close to the star where the shear is high.} 


\subsection{Caveats and Future Work}
\label{sec:caveats}
In this work, we have omitted two potentially important effects. The first is that massive cores are magnetized and therefore magnetic pressure may be dynamically important during core collapse. The second is that we neglect other feedback mechanisms such as magnetically launched collimated outflows and photoionization, which will also affect the growth rate of massive stars and core fragmentation. We also note that simulations presented in this work focus on the collapse of very dense massive star-forming cores ($\Sigma = 1 \; \rm{g \; cm^{-3}}$) and are therefore more representative of cores that sample the high density end of massive star formation since massive stars form in cores and clumps with surface densities that range from $\Sigma \sim 0.1-1 \; \rm{g \; cm^{-3}}$ \citep{Krumholz2007d, Tan2014a}. Hence, \add{our choice to neglect magnetic fields and} chosen initial conditions are biased towards faster collapse time scales and higher accretion rates than those typically observed in most galactic massive star forming regions that have accretions rates of $\sim 10^{-5} - 10^{-4} \; M_{\rm \odot} \; \rm{yr^{-1}}$  and collapse time scales of $\sim 10^5$ yr \add{thereby yielding higher star formation efficiencies (SFEs) per free-fall time than those typically observed} \citep{Zhang2015a}. 

Magnetic support in cores can slow down their collapse, reduce fragmentation, and result in lower accretion rates onto stars \add{thereby reducing the SFE of massive pre-stellar cores} \citep{Myers2014a, Burkhart2015a}. Observations of molecular clouds find that they are super-critical ($\mu_{\phi} \gtrsim 1$ where $\mu_{\phi}$ is the mass-to-flux ratio) concluding that gravity dominates over magnetic pressure in molecular clouds \citep{Troland2008a, Crutcher2010a}. In addition, most observed massive dense cores are highly super-critical ($\mu_{\phi} \gg 1$). For example, \citet{Girart2013a} studied the evolved massive dense core DR 21(OH) and found $\mu_{\phi} \approx 6$ whereas its parent filament has a lower value of $\mu_{\phi} \approx 3.4$. Additionally, \citet{Ching2017a} studied several more cores in this filament and found that all cores are roughly or highly super-critical with $\mu_{\phi} \approx 1-4.3$). Hence, their results suggest that as filaments and clumps collapse to form dense cores gravity can become even more important making magnetic regulated collapse subdominant. 

In addition to gravity, turbulence can also dominate over magnetic pressure in massive cores and clumps. Both numerical simulations and observations have shown that on large scales (i.e., $l \gtrsim 0.1-100$ pc) the ISM is largely sub-Alfv\'enic whereas on smaller size scales, such as the size scales of dense cores (e.g., $0.01 \lesssim l  \lesssim 0.1$ pc), molecular gas becomes super-Alfv\'enic with typical values of $\mathcal{M}_{\rm A} \approx 2-3$ suggesting that turbulence may dominate over magnetic fields in most massive star forming cores \citep{Hull2017a, Zhang2019a, Hull2019a}. However, there are a few cases in which the massive core is super-critical but sub-Alfv\'enic. For example, \citet{LiuLi2018a} studied the massive c8 core ($\sim200 \; M_{\rm \odot}$) in G035.39-00.33 and found that the magnetic field structure of c8 has an hour glass morphology that arises from the field lines being dragged in as the core forms due to filament collapse. This effect increases the magnetic pressure in the inner core causing it to have an infall velocity smaller than the core envelope. Likewise, \citet{Zhang2014a}  found that the magnetic field orientation in massive star forming regions ranges from ordered hour-glass configurations likely owing to strong magnetic fields to more chaotic distributions that are a result of weak magnetic fields. They also found that the magnetic field throughout the clump shows preferential alignment and aligns from the clump to the core scale. Their results suggest that at least qualitatively the field may be dynamically important. However, a more complete statistical sample is required to determine if cores are preferentially sub- or super-Alfv\'enic \citep{Hull2019a}. Strong magnetic fields will also affect clump fragmentation and protocluster formation. \citet{Fontani2018a} concluded that cores that show fragments distributed in a filamentary-like structure are likely characterized by a strong magnetic field. Hence, future work requires including magnetic fields and simulating the collapse of sub- and super-Alfv\'enic cores to determine how the magnetic field and its relative importance to gravity and turbulence affects the growth rate of massive stars and the fragmentation properties of collapsing massive dense cores.


The simulations presented in this work also neglected how feedback from collimated outflows and photoionization affects the accretion history of massive stars and fragmentation of massive cores. Outflows are magnetically launched and present during the accretion phase for both low-mass and high-mass star formation with typical mass-loss rates of $\sim10-30$\% of the accretion rate \citep{Arce2007a, Carrasco-Gonzalez2010a, Carrasco-Gonzalez2015a, Maud2015a}. The presence of outflows will reduce the growth history of accreting massive stars \add{yielding lower SFEs  for massive pre-stellar cores} and therefore reduce the evolutionary effect of radiative heating in the collapsing core since the presence of outflows provides an avenue for the the radiative flux to escape \citep{Cunningham2011a}. Such an effect will lead to a higher degree of thermal fragmentation, potentially resulting in more companion stars to the massive star regardless of the initial virial state of the pre-stellar core.

Photoionization, on the other hand, will only become important when the star is sufficiently massive and compact enough to generate a large amount of photoionizing radiation to produce an ultra compact HII region and this occurs at $M_{\rm \star} \sim $25-35 $M_{\rm \odot}$ \citep{Hosokawa2009a, Kuiper2018a}. Recent RHD simulations of massive star formation by \citet{Kuiper2018a} that includes radiation pressure, outflows, and photoionization found that the larger thermal pressure from the resulting ionized gas ($T = 10^4$ K) can actually enhance accretion onto the star when photoionization first becomes important. They also show that photoionization feedback broadens the outflow cavities but does not limit disk accretion onto the star. Instead, they conclude that radiation feedback \add{on dust} at late times is responsible for reducing the gravitational infall of core material onto the circumstellar disk and eventually cuts off accretion. Hence, outflow and photoionization feedback likely does not shut off mass accretion onto massive stars but these feedback components will aid in reducing the accretion rate onto massive stars.

\addd{Finally, we note that modeling the collapsing cores as isolated objects is an oversimplification since observed massive cores and clumps are typically found in embedded, dynamical environments that may feed the cores' material as they collapse \citep[e.g.,][]{Avison2019a}. This type of environment may affect the fragmentation and collapse of such cores. However,} the simulations presented in this work provide a crucial first step in understanding how the dynamic state of massive pre-stellar cores and stellar feedback affects the growth rate of massive stars and core fragmentation. In future work we will also address the concerns presented above by including magnetic fields and feedback from protostellar outflows in addition to radiation pressure to further understand how the dynamic state of the pre-stellar core affects the formation history and multiplicity of massive stars. 

\section{Conclusions}
\label{sec:conc}
In this work we performed 3D RHD simulations of the collapse of turbulent massive pre-stellar cores that are identical in every way except for their initial virial state to determine how the dynamic state of massive pre-stellar cores affects the formation history of massive stars, core fragmentation, and the multiplicity of massive stellar systems. To address these questions we modeled the collapse of a highly subvirial core with $\alpha_{\rm vir} = 0.14$ and a roughly virialized core with $\alpha_{\rm vir} = 1.1$. Our simulations included radiation feedback from both the direct stellar and dust-reprocessed radiation fields inherent to massive star formation owing to the massive star's large luminosity as it is actively accreting. We note that the simulations presented in this work neglect magnetic fields and outflow feedback that are ubiquitous in massive star forming regions. However, we will address these physical processes in future work to determine how these processes further affect the growth of massive stars and fragmentation of massive cores.

Our main results are summarized as follows. We find that subvirial cores undergo a fast global collapse leading to higher accretion rates, by up to a factor of $\sim2$, onto the massive star \addd{as compared to the accretion rates obtained from the gradual inside-out collapse of virialized cores.}
Additionally, we find that virialized cores undergo significant turbulent fragmentation into companion stars at early times owing to their greater degree of turbulence (i.e., higher $\mathcal{M}$). The faster growth rate of massive stars that originate from less turbulent, subvirial cores leads to higher stellar luminosities at earlier times and therefore enhances the radiative heating from both the direct stellar and indirect dust-reprocessed radiation fields, thereby suppressing turbulent fragmentation. We also find that the global gravitational collapse of subvirial cores and gradual inside out collapse of virialized cores increases the velocity dispersion of massive cores in agreement with observations \citep{Traficante2018a}. Furthermore, radiative feedback from massive stars also increases the averaged mass-weighted velocity dispersion and temperature of massive cores.  

Regardless of the core's initial virial state, we find that an optically thick accretion disk forms around the massive star and supplies material to the star, especially at late times when radiation pressure has driven radiation pressure dominated bubbles above and below the accretion disk. We also note that the accretion disk forms around the massive star earlier for virialized initial conditions because of  the core's larger angular momentum content and the disk is much larger in size at late times as compared to the accretion disk that forms around the massive star in our subvirial run. At late times, the accretion disk undergoes disk fragmentation forming close-in low mass companions \add{regardless of the core's initial virial state yielding multiple systems.} 

\software{\textsc{yt}, \citep{Turk2011a}, \orion\ \citep{Li2012a}, \harm\ \citep{Rosen2017a}}

\subsection*{Acknowledgements}
The authors thank the anonymous referee for their advice and suggestions which greatly improved  the manuscript. A.L.R. acknowledges support from NASA through Einstein Postdoctoral Fellowship grant number PF7-180166 awarded by the \textit{Chandra} X-ray Center, which is operated by the Smithsonian Astrophysical Observatory for NASA under contract NAS8-03060. PSL acknowledges support by NASA through a NASA ATP grant NNX17AK39G. B.B. acknowledges support from the Simons Foundation. A.L.R would like to thank Alyssa Goodman, Phil Myers, and Alessio Traficante for insightful conversations regarding this work.
\bibliographystyle{apj}
\bibliography{refs}

\end{document}